\documentclass[12pt]{JHEP3}

\usepackage{amsmath}
\usepackage{amssymb}
\usepackage{graphicx}

\newcommand{\be}{\begin{eqnarray}}
\newcommand{\ee}{\end{eqnarray}}
\newcommand{\nn}{\nonumber}
\newcommand{\bn}{\begin{enumerate}}
\newcommand{\en}{\end{enumerate}}

\def\IC{\mathbb{C}}

\def\IZ{\mathbb{Z}}

\def\CM{{\cal M}}
\def\CN{{\cal N}}

\def\CW{{\cal W}}

\def\a{\alpha}

\def\d{\delta}
\def\e{\epsilon}

\def\i{\iota}

\def\l{\lambda}
\def\m{\mu}
\def\n{\nu}

\def\Tr{{\rm Tr}}

\def\det{{\rm det}}

\title{  ABCD of  3d ${\cal N}=8$ and $4$ \\
Superconformal Field Theories }

\author{ Dongmin Gang$^1$, Eunkyung Koh$^1$, Kimyeong Lee$^1$ and  Jaemo Park$^{2}$

\\
\\
$^1$Korea Institute for Advanced Study, Seoul 130-012, Korea
\\
$^2$Department of Physics \& PCTP, POSTECH,
Pohang 790-784, Korea
\\
\\
E-mail: \email{ arima275, ekoh, klee@kias.re.kr, jaemo@postech.ac.kr} } 

\abstract{We argue the equivalence between the infrared conformal
field theory of the 3d $\CN=8$ supersymmetric Yang-Mills theories of
ABCD ($U(N), SO(2N+1), Sp(2N), O(2N)$)  gauge groups and the ABJ(M)
theories of $U(N)_k\times U(\tilde N)_{-k}$ for $k=1,2$. We support
this duality by comparing  the superconformal index of the IR limit
of these super Yang-Mills theories and that of those ABJ(M) models.
Especially   we find the match between two indices of  (mirror dual
of) the $\CN=8$ $U(N)$ SYM and of $U(N)_1\times U(N)_{-1}$ ABJM
model. Also we take large $N$ limit of ABCD  super Yang-Mills
theories with additional fundamental hyper-multiplets and infer the
large N limit of $\CN=8$ ABCD theories themselves, finding the
expected gravitational duals. With the additional input on finite N,
we argue the equivalence of Yang-Mills and ABJ(M) theories for all
N.  We further explore similar dualities to Chern-Simons matter
theories for $\CN=4$ Yang-Mills theories related by mirror symmetry.
}

\preprint{KIAS-P11031}

\begin{document}

\section{Introduction}

Recently one  have witnessed rapid progress in the understanding of
the superconformal field theories (SCFT) associated with M2 branes.
The famous example is the theory on $N$ M2 branes on $\IC^4$, which
is  realized as the ABJM model which is a ${\cal N}=6$ Chern-Simons
matter theory with gauge group $U(N)_k \times U(N)_{-k}$ and
Chern-Simons level $k=1$~\cite{ABJM}.   Prior to this development,
strongly coupled  superconformal field theories(SCFT) have been
explored as the infrared (IR) limit of supersymmetric Yang-Mills
(SYM) theories in three
dimensions~\cite{Seiberg,Sethi,Berkooz:1998sn}. The recent progress
has shown especially that the IR limit of the ${\cal N}=8$ $U(N)$
SYM theory flows to the ABJM model with $U(N)_1\times U(N)_{-1}$.
Especially the calculation of the partition functions of two
theories matches each other beautifully~\cite{Kapustin:2010xq}.

An obvious question is whether there is any similar superconformal Chern-Simons matter theory
which describes the IR limit, or the infinite coupling limit, of ${\cal N}=8$ super Yang-Mills
theory with the general gauge group $G$.   If that is the case, one  would like to test   the equivalence or duality.

In this work, we explore  the IR limit of the $\CN=8$ SYM theories
of the gauge group in the ABCD classical groups, say, $U(N),
SO(2N+1), Sp(2N), O(2N)$ . We argue that the   $ O(2N)$ Yang-Mills
theory flows to the $U(N)_2\times U(N)_{-2}$ ABJM model
 and the    $SO(2N+1) $ Yang-Mills theory flows to
$U(N)_2\times U(N+1)_{-2}$ ABJ theory~\cite{Aharony:2008gk}. The $\CN=8$ $Sp(2N)$ SYM theory can  flow to either ones.
 The main tool we use  is the superconformal index~\cite{Bhattacharya:2008zy,Kim:2009wb,Imamura:2011su}.

 There are four kinds of orientifold planes $O2^-, O2^+, \tilde O2^+,\tilde O2^-$ in type IIA
 string theory and two kinds of  orbifold plane $OM2^-$ and $OM2^+$ in the M theory.
 Our  BCD $\CN=8$ supersymmetric Yang-Mills theories originates from the theory on
 D2 branes near the orientifold planes. Each orientifold plane is made of
 two $OM2$ planes at the end of 11d line segment   in the M-theory.
 We understand the $\CN=8$ super Yang-Mills theory on D2 branes near the
 orientifold and the ABJ(M) model near $OM2$ plane. This naturally leads to the
 correspondence between super Yang-Mills theories and ABJ(M) models.

Our main tool to test the equivalence is the superconformal index.
  However  the usual localization is not directly applicable to the index
computation of  ${\cal N}=8$ $U(N), O(2N),SO(2N+1), Sp(2N)$
SYM theories. For the $U(N)$ case, we use the $\CN=4$ mirror dual, or the $\CN=8$ theory
with an additional fundamental hyper-multiplet,  which is similar to the one used
for the calculation of the partition function~\cite{Kapustin:2010xq}. The match
between the index for the $\CN=8$ $U(N)$ SYM theory and that of $U(N)_1\times U(N)_{-1}$
is tested explicitly in small $N$ in the series expansion. For the $O(2N),SO(2N+1), Sp(2N)$ cases,
the IR SCFTs of the $\CN=8$ SYM theories are not equivalent to the IR SCFT limits of
the $\CN=8$ SYM theories with additional  fundamental hyper-multiplet.

Thus, we take the roundabout approach to the   BCD cases. For small
$N$, we improvise and obtain the indices and show that they are
identical to those of ABJ(M) type with $k=2$.  In addition to
it, we
 include arbitrary $m$ fundamental hyper-multiplets to  super Yang-Mills
theories. The corresponding field theoretic index is calculable and
we consider its  large $N$ limit. This has the gravitational dual as
the orbifold of $AdS_4\times S^7/\IZ_2$ of order $m$ whose index has
the contributions from the twisted sector. After subtracting off the
contribution of the twisted sectors from the field theory index, one
obtains $\IZ_m$ invariant sectors of $AdS_4\times S^7/\IZ_2$ out of
the $\CN=8$ super Yang-Mills theory. By varying $m$, one can see
that large $N$ limit of $\CN=8$ BCD super Yang Mills theory matches
with the gravitational calculation on $AdS_4\times S^7/\IZ_2$ space.
This in turn can be identified with the index of the ABJ(M) theory
with $k=2$ as some calculations was done for this AdS/CFT
correspondence between the gravitational calculation and the ABJ(M)
field theoretic calculation has been tested before in
Ref.~\cite{Kim:2009wb}.
 By further working out the particular orbifold theories for finite $N$ or
by considering the Higgsing down to small $N$ theories,
 one could  confirm the proposed dualities for all $N$.

Once handling ${\cal N}=8$ cases, we can ask the similar
question for ${\cal N}=4$ Yang-Mills theories. Here our main focus is the
theories arising in the mirror symmetry. As is well known under the
mirror symmetry the Coulomb   and Higgs branches  are interchanged. The $\CN=4$ superconformal field theory is living
at the origin of the moduli space where Coulomb and Higgs branch
intersecting.   One can
ask if such SCFT can be described again in terms of supersymmetric Chern-Simons matter theory.
 We mainly consider ${\cal N}=4$
Yang-Mills type theories which is describable by Hanany-Witten set-up
with D3/NS5/D5 branes~\cite{Hanany:1996ie}. As is well known, the mirror symmetry is
realized as S-dual transformation in the Hanany-Witten setup.
Interestingly the associated superconformal Chern-Simons matter  theory can be obtained by
 T-dual transformation where $\tau\rightarrow \tau+1$ with $\tau$ being axion-dilaton of
 Type IIB theory. We carry out the index
computations, which impressively confirm  our proposal on the ${\cal N}=4$ SCFTs.

The contents of our paper is as follows. In Sec.2,  we briefly
review   orientifold planes which made   of $OM2$ planes, and
present the proposal for the IR limit of the $\CN=8$ supersymmetric
Yang-Mills theories in terms of the ABJ(M) models with $k=1,2$. In
Sec.3, we compute the index of $U(N)_1\times U(N)_{-1}$ ABJM theory
and that of the   mirror dual of ${\cal N}=8$ $U(N)$ SYM, and  find
the prefect agreement. And we carry out the detailed exercise of the
large N dual of ${\cal N}=8$ $ U(N)$ SYM by working out the field
theory index of orbifolded theories of ${\cal N}=8$ $ U(N)$
Yang-Mills theory and subtracting the twisted sector contribution in
the gravitational side.  In Sec. 4, we carry out the similar
analysis for ${\cal N}=8$ $ O(2N), SO(2N+1), Sp(N)$ theory and show
that their gravitational dual is $AdS_4 \times S^7/{\mathbb Z}_2$ in
the large N limit. With additional input about the index for small
$N$ or by consideration of Higgsing pattern, one can match
Yang-Mills theories to ABJ(M) theories with Chern-Simons level $2$.
 In Sec. 5, we work out various super Chern-Simons matter(SCSM) realizations associated with
${\cal N}=4$ SCFT appearing in the ${\cal N}=4$ mirror symmetry.
Again we use the index computation as a main tool to confirm the
mirror pair and the associated ${\cal N}=4$ SCFT as SCSM. We also
consider the partition function with mass and Fayet-Iliopoulos terms
and work out how such parameters are mapped under the duality
between ${\cal N}=4$ super-Yang-Mills and SCSM. Various technical
details are relegated to several appendices.

\section{  3d $\CN=8$ Susy Yang-Mills Theories}

The 3d ${\cal N}=8$ susy Yang-Mills theory is obtained from the dimensional reduction of ${\cal N}=1$
10d super Yang-Mills theory. The bosonic part of the action is
given by
\begin{equation}
 \frac{1}{g^2}\int d^3x \, {\rm Tr} \Big(-\frac14 F_{\mu\nu}^2-\frac12\sum_{i=1}^7(D_{\mu}
\phi_i)^2+\frac14 \sum_{i,j=1}^7([\phi_i,\phi_j])^2 +\frac12
\e^{\mu\nu\rho} D_\rho( F_{\mu\nu}\vartheta) \Big).
\label{basicaction}
\end{equation}
The last term affects the dynamics as the expectation value of the dual scalar fields of the gauge fields is
fixed by the parameter $\vartheta$.  Let us first consider the theory with the gauge group $U(N)$. Along the
moduli space all $A_{\mu}, \phi_i$ are commuting so that the gauge group is broken to
$U(1)^N$. Along the flat directions, we have
\begin{equation}
\sum_{a=1}^N \Big(  -\frac1 {4g^2} F^a_{\mu\nu}F^{a\mu\nu}
-\frac1{2g^2} \sum_{i=1}^7\partial_\mu \phi_i^a \partial^\mu\phi_i^a+  \epsilon^{\mu\nu\rho} F^a_{\mu\nu}\partial_\rho \phi_8^a \Big)\;,
\end{equation}
where each element belongs to Cartan torus of $U(N)=U(1)^N$. Here we rescaled compact scalars to have $2 \pi$ periodicity ($\phi_8^a \sim \phi_8^a + 2\pi$)
\begin{align}
\vartheta = \sum_{a=1}^N 2 g^2 \phi_8^a t^a  \;, \label{dual photon}
\end{align}
where $t^a$s are generators of unbroken $U(1)^N$ gauge group with normalization $\Tr(t^at^b) = \delta^{ab}$.
One can dualize the photon with field equation $F^{a \mu\nu}/g^2= 2\e^{\m\n\rho}\partial_\rho \phi^a_8$ and gets
\begin{equation}
-\frac12 \sum_{a }\int d^3x \Big( \sum_{i=1}^7\frac{1}{g^2}(\partial \phi_i^a)^2+4 g^2 (\partial_\mu \phi^a_8)^2 \Big)
\end{equation}
Thus the moduli space for the $\CN=8$ susy Yang-Mills theory with the gauge group $A_{N-1}=U(N)$ is \cite{Seiberg}
\begin{equation}
\CM= \frac{({\mathbb R}^7\times S^1)^N}{S_N}   \label{moduli}
\end{equation}
where $S_N$ is the permutation group of $N$ elements. In the IR
limit, $g\rightarrow \infty$ and the radius of the circle associated
with the compact scalar tends to infinite. In this case the moduli
space is given by
\begin{equation}
\CM_{U(N)}=\frac{{\mathbb C}^{4N}  }{S_N}. \label{modun}
\end{equation}
Note that $S_N$ is the Weyl group of $U(N)$ gauge group. More
generally for $N=8$ Yang-Mills theory with gauge group $G$ of rank $r$, the moduli space
is supposed to be
\be \CM= \frac{ {\mathbb R}^{7r }\times \hat T^r }{\CW_G}
\label{modun2} \ee
where $\CW_G$ is the Weyl group of $G$ and $\hat{T}^r$ is the Cartan torus for the dual group~\cite{Seiberg}.
In the IR limit the moduli space is given by
\begin{equation}
\CM_G=\frac{{\mathbb C}^{4r}}{\CW_G}
\end{equation}

For $B_N=SO(2N+1), C_N= Sp(2N), D_N=O(2N)$ the Weyl group is $\IZ_2^N\times
S_N$ so that the moduli space for all these gauge groups~\cite{Berkooz:1998sn} is
\begin{equation}
\CM_{BCD}= \frac{({\mathbb C}^4/{\mathbb Z}_2)^N}{S_N}.  \label{m2}
\end{equation}
Interestingly enough, the vacuum moduli space of $SU(N)$ and
$SO(2N)$ are more complicated. For the $SU(N)$ case, the vacuum
moduli space is
\be \CM_{SU(N)} =\frac{{\mathbb C}^{4(N-1)}}{S_N} \ee
and the $SO(2N)$ has   the Weyl group $\CW_{SU(N)}={\mathbb
Z}_2^{N-1} \times S_N$ so that its moduli space is
\be \CM_{SO(2N)}= \frac{({\mathbb C}^4/\IZ_2)^{ N}    }{  S_N} \times \IZ_2
\ee

One   notices the   vacuum moduli space of the low energy limit or infinite coupling limit of
the given $\CN=8$ super Yang-Mills theory becomes simpler for $U(N), SO(2N+1), Sp(2N), O(N)$
gauge groups. Essentially, they allow the interpretation in term of   multiple M2 branes exploring
either ${\mathbb C}^4 $ or the orbifold  ${\mathbb C}^4/{\mathbb Z}_2$.

To understand this view, let us first consider the dynamics of $N$
 parallel D2 branes of type IIA string theory in the flat space-time. The transverse 7d space
 can be either flat ${\mathbb R}^7$ or ${\mathbb R}^7/{\mathbb Z}_2$ orbifold.
 In type IIA string theory, there are four kinds of orientifolds: $O2^{-}, O2^+,\widetilde{O2}^+,
 \widetilde{O2}^-$ of D2 charge, $-1/8,+1/8, +1/8,  +3/8$, respectively. The $\CN=8$ super
 Yang-Mills theories on $N$ D2 branes in the background of these orientifold have the gauge group
 $O(2N), Sp(2N), Sp(2N), SO(2N+1), $ respectively. The gauge group $Sp(2N)$ arises for the two cases
 with different range of diagonal $Sp(2N)$ matrix $\vartheta$ whose eigenvalues denote the positions of
 M2 branes in the $x^{11}$ direction. Table I below denotes the $O2$ planes which are composed of
 two $OM2$ planes at two ends of the compact line segment $x^{11}$ of M-theory.

\begin{center}
\begin{tabular}{|c|c|c|}
\hline
$O2^-$ & $OM2^-+OM2^- $ & $O(2N)$ \\
$O2^+, \widetilde{O2}^+$ & $OM2^- + OM2^+ $ & $Sp(2N)$ \\
$\widetilde{O2}^-$ & $OM2^+ + OM2^+$ & $SO(2N+1)$ \\
\hline \end{tabular}
\vskip 0.5cm
\centerline{Table I:   $O2$ Plane Made of Two $OM2$ Planes}
\end{center}

In the M-theory $\IC^4/{\mathbb Z}_2$ orbifold singularity can come in two varieties depending
on the presence of discrete torsion. Without discrete torsion, it is called $OM2^{-}$ plane of M2
charge $-1/16$ and with a discrete torsion, it becomes $OM2^+$ plane with quarter of M2 brane stuck
with  M2 brane charge $+3/16$~\cite{Berkooz:1998sn}. A M2 brane on all of these orientifold background
has the same moduli space $\IC^4/\IZ_2$. The superconformal field theory on $N$ M2 branes exploring $\IC^4 $
is the $\CN=6$ ABJM model of the gauge group $U(N)_1\times U(N)_1$ with $k=1$, whose supersymmetry gets
enhanced to $\CN=8$~\cite{susy}. The theory   near $OM2^-$ is that of the gauge group $U(N)_2\times U(N)_{-2}$
with $k=2$, whose supersymmetry is also enhanced to  $\CN=8$.  The theory near $OM2^+$ is  that of the gauge
group $U(N)_2\times U(N+1)_{-2}$ with $k=2$, whose supersymmetry is also enhanced to $\CN=8$~\cite{Bashkirov:2011pt}.

Let us first note that there exists an   important  equivalence in physics or duality between
the ABJ(M) models~\cite{Aharony:2008gk}:
\be U(N +\ell)_k\times U(  N)_{-k} \Longleftrightarrow U(N)_k\times U(N+k-\ell)_{-k} .
\ee
This duality implies the following duality between three $\CN=8$ models for $k=1$:
\be U(N+1)_1\times U(N)_{-1} \Longleftrightarrow
U(N)_1\times U(N)_{-1}   \Longleftrightarrow U(N)_1\times U(N+1)_{-1} \ . \ee
For $k=2$, the the following duality holds:
\be U(N+2)_2\times U(N)_{-2} \Longleftrightarrow  U(N)_2\times U(N)_{-2}
 \Longleftrightarrow  U(N)_2\times U(N+2)_{-2} \ .\ee
One can confirm this duality by calculating the superconformal indices of the next section and comparing them.
For example, we find the exact match for $N=1$ case. The  duality
\be U(N+1)_2\times U(N)_{-2} \Longleftrightarrow  U(N)_2\times U(N+1)_{-2}
\ee
implies that the model of this gauge group is parity even. The
parity for this model is given by the usual parity transformation
accompanied by the Seiberg-like dualities \cite{Bashkirov:2011pt}.

Let us now compare the SCFTs limit of the $\CN=8$ supersymmetric
Yang-Mills theories with the ABJM and ABJ type superconformal
Chern-Simons matter theories with $U(N)_k \times U(N+1)_{-k}$ for
$k=1,2$. From the brane picture it is obvious now. First of all the
vacuum moduli space of the IR limit of the super Yang-Mills theory
and that of ABJ(M) should match. The number of supersymmetry should
be $\CN=8$. The Table II shows the relation between the IR limit of
$\CN=8$ super Yang-Mills theory and the ABJ(M) models.

\begin{center}
\begin{tabular}{|c|r|l|}
\hline
Type & Super Yang-Mills & Super Chern-Simons  \\
\hline
A & $U(N)$ SYM &  $U(N)_1\times U(N)_{-1}$   \\
B & $SO(2N+1)$ SYM & $U(N)_2\times U(N+1)_{-2}$   \\
C & $Sp(2N)$ SYM & $ U(N)_2\times U(N+1)_{-2}$  \\
C & $Sp(2N)$ SYM & $ U(N)_2\times U(N)_{-2}$  \\
D & $O(2N) $ SYM  & $U(N)_2\times U(N)_{-2} $  \\
\hline
\end{tabular}
\vskip 0.5cm
\centerline{Table II: The IR limit of Super Yang-Mills Theories as ABJ(M) Models}
\end{center}

As we suspect the equivalence between the IR limit of super Yang-Mills theory and the ABJ(M) models,
we expect many quantities of two theories should match.   Especially the partition functions on $S^3$
and the superconformal indices should be identical.   In the subsequent
sections, we explore this equivalence or duality between these theories by calculating their indices.
In the appendix C, we attempt to calculate the partition function and see some matches also.

\section{The index of $\CN=8$ $U(N)$ Super Yang-Mills Theory}

The 3d $\CN = 8$ $U(N)$ super Yang-Mills theory arises as the field
theory dynamics on $N$ D2 branes. As the IR limit, the strong
coupling limit of the theory on D2 branes becomes the theory on $M2$
branes, which has been identified with the  ABJM model with
$U(N)_k\times U(N)_{-k}$ and $k=1$.  The index calculation for the
ABJM model for all $N,k$ has been done in Ref.~\cite{Kim:2009wb}.
What is new here is   the index for the $\CN=8$ super Yang-Mills
theory. In this section, we test the equivalence in terms of the
index.

The equivalence between the IR limit of super Yang-Mills theory and
the ABJM theory has been well tested in terms of the partition
function~\cite{Kapustin:2010xq}. Similar to the partition function,
there is the issue of the divergence which can be avoided similarly
by considering the mirror dual which is $\CN=4$ super Yang-Mills
theory with one adjoint and one fundamental hyper-multiplets. We
also approach the index in the large $N$ limit by the field theory
and gravity. By considering the Higgsing pattern, we can also see
the consistency of the proposed duality between  $\CN = 8$ $U(N)$
super Yang-Mills theory and the ABJM theory with Chern-Simons level
1.

\subsection{Modules of the Index} \label{generalindex}

Let us start by discussing   the general structures of the index for the 3d $\CN=2$ superconformal field
theories (SCFTs).
Superconformal index for higher supersymmetric theory
can be defined using their $\mathcal{N}=2$ subalgebra. The bosonic
subalgebra of the 3-d $\mathcal{N}=2$ superconformal algebra is
$SO(2,3) \times SO(2) $.  There are three Cartan elements denoted by
$\epsilon, j_3$ and $R$ which come from three factors
$SO(2)_\epsilon \times SO(3)_{j_3}\times SO(2)_R $ in the bosonic
subalgebra.  One can define the superconformal index for 3-d
$\mathcal{N}=2$ SCFT as follows \cite{Bhattacharya:2008zy},
\begin{equation}
I=\Tr (-1)^F \exp (-\beta'\{Q, S\}) x^{\epsilon+j_3},
\label{def:index}
\end{equation}
where $Q$ is a special  supercharge with quantum numbers
$\epsilon = \frac{1}2, j_3 = -\frac{1}{2}$ and $R=1$ and $S= Q^\dagger$.
They satisfy following anti-commutation relation:
\begin{equation}
 \{Q, S\}=\epsilon-R-j_3  .
\end{equation}
In the index formula, the trace is taken over gauge-invariant local
operators in the SCFT defined on $\mathbb{R}^{1,2}$ or over states
in the SCFT on $\mathbb{R}\times S^2$. As is usual for Witten index
\cite{Witten:1981nf}, only BPS states satisfying the bound $\epsilon-R-j_3
=0 $ contributes to the index and the index is independent of
$\beta'$. If we have additional conserved charges commuting with
chosen supercharges ($Q,S$), we can turn on the associated chemical
potentials and the index counts the number of BPS states with the
specified quantum number of the conserved charges.

The  superconformal index is exactly calculable using localization
technique \cite{Kim:2009wb},\cite{Imamura:2011su}. Following their
works, the superconformal index can be written in the following form
(for simplicity, we turn off the chemical potentials except for $x$)
\begin{align}
I(x) = \sum_{\{s\}} \int d \sigma \; x^{\epsilon_0} \exp [i S_0  ]
\exp \big{[} \sum_{p=1}^\infty \frac{1}p f_{tot} (x^p, e^{i p
\sigma_i} )  \big{]}. \label{index structure}
\end{align}
We are considering 3-d $\mathcal{N}\geq 4$ super symmetric
Chern-Simons matter theory (SCSM) with gauge group $G$ and
hyper-multiplets in $R_I$ (chiral-multiplets in $R_I$ and
$\bar{R}_I$) of $G$. To take trace over Hilbert-space on $S^2$, we
impose proper periodic boundary conditions on time direction
$\mathbb{R}$. As a result, the base manifold become $S^1\times S^2$.
For saddle points in localization procedure, we need to turn on
monopole fluxes on $S^2$ and holonomy along $S^1$. These
configurations of the gauge fields are denoted by  $\{s\}$ and $\{
\sigma \}$ collectively. Both variables take values in the Cartan
subalgebra of $G$. $S_0$ denote the classical action for the
(monopole+holonomoy) configuration on $S^1\times S^2$. $\epsilon_0$
is called the Casimir energy,
\begin{equation}
\epsilon_0 = \frac{1}2 \sum_{I}\sum_{\rho \in R_I}  |\rho(s)| -
\frac{1}2 \sum_{\alpha \in G} |\rho(s)|.
\end{equation}
Here $\rho \in R_I$ represent the weights of representation $R_I$
and $\alpha \in G$ denote the roots (weight of adjoint) of $G$. The
$f_{tot}$ can be divided into two parts, $f_{tot} =
f_{hyper}+f_{vec}$ where
\begin{eqnarray}
&& f_{hyper}(x, e^{i \sigma}) =\sum_{I} \sum_{\rho \in R_I}
\frac{x^{1/2}}{1+x} x^{|\rho(s)|} \big{[} e^{i \rho(\sigma)} + e^{-
i \rho(\sigma)} \big{]}, \nonumber
\\
&& f_{vec}(x, e^{i \sigma}) = -\sum_{\alpha \in G ; \alpha \neq 0 }
e^{i \alpha (\sigma)}x^{|\alpha(s)|},
\end{eqnarray}
 where we used the fact that the conformal dimensions of fields in
hyper-multipet is canonically $1/2$  for 3-d ${ \cal N} \geq 3$ theories.
The chiral superfields in $\mathcal{N}=4$ vector-multiplet have
conformal dimension $1$ and does not contribute to the index. If the
action contains the Chern-Simons terms, it gives the nonvanishing
contribution,
\begin{equation}
i S_0=\frac{i k}{4\pi}\int tr(A_0\wedge   dA_0-\frac{2i}{3}A_0\wedge
A_0 \wedge A_0)=i k tr (\sigma s)
\end{equation}
 where $k$ is the Chern-Simons level.

For super Yang-Mills, we are  taking  the IR limit
$g_{YM}\rightarrow \infty$ in the above index formula. This makes
$S_0 =0$ in this case. This strategy works for many cases for the
computation of the partition function of the Yang-Mills theories
with ${\cal N}=4$ supersymmetry. One criterion for the existence of
the smooth limit is that $SO(4)_R$ symmetry appearing in the
Lagrangian of super Yang-Mills theory becomes a part of the
superconformal symmetry. If this holds, the scalars in the vector
multiplets and the gauge fields have the IR conformal dimension 1
and becomes irrelevant so that we can drop the kinetic term of the
gauge fields \cite{Kapustin:2010xq}. For all of the Yang-Mills
theories whose index is computed in this paper, this criterion is
satisfied.

\subsection{the Index of ABJ(M) Model}

 For the ABJM (ABJ) model with $U(N)_k\times U(\tilde N)_{-k}$, let us re-derive the index
 result~\cite{Kim:2009wb}. This is supersymmetric Chern-Simons matter theory with the gauge group
 $U(N)_k\times U(\tilde N)_{-k}$ with the subscript denoting the
Chern-Simons level of the underlying gauge group. The matters
consist of two hypermultiplets $A,B$ in $(N,\bar{\tilde N})$ and $(\bar{N},\tilde N)$ of
$G$,
\begin{align}
\begin{array}{ccc}
    & U(N)_k & U(\tilde N)_{-k} \\
  (A_1 , B_1) & N & \bar{\tilde N} \\
  (A_2 , B_2) & \bar{N} & \tilde N.
\end{array}
\end{align}
$(A_1,A_2)$ and $(B_1,B_2)$ denote the chiral multiplets in $A$ and $B$.

Monopole charges are denoted by $\{s \}=\{n_i, \tilde{n}_{\tilde \j}
\}$   and holonomy variables by $\{\sigma\} = \{
\lambda_i, \tilde{\lambda}_{\tilde \j}\}$, where
$i=1,\ldots, N$ and $ \tilde \j =1,\ldots,\tilde N $. Weights are given
by
\begin{align}
&\rho \in (N, \bar{\tilde N}) \; : \; \{ e_i - \tilde{e}_{\tilde \j} \}, \nonumber
\\
&\rho \in (\bar{N}, \tilde N) \; : \; \{ -e_i +\tilde{e}_{\tilde \j} \}, \nonumber
\\
& \alpha \in G \; : \; \{ e_i - e_j, \tilde{e}_{\tilde \i} - \tilde{e}_{\tilde \j} \}.
\end{align}
The terms appearing in the   general formula (\ref{index structure})
are given as
\begin{align}
&S_0 = i k \sum_{i=1}^N  n_i \lambda_i - ik\sum_{{\tilde \i}=1}^{\tilde N}\tilde{n}_{\tilde \i}
 \tilde{\lambda}_{\tilde \i} ,
\nonumber
\\
&f_{hyper}(x, e^{i \lambda}, e^{i \tilde{\lambda}}) =2
\sum_{i,\tilde{\j} }  \big{(}  \frac{x^{1/2}}{1+x} x^{|n_i- \tilde{n}_{\tilde \j}|}
e^{i (\lambda_i - \tilde{\lambda}_{\tilde \j})} +  \frac{x^{1/2}}{1+x}
x^{|n_i- \tilde{n}_{\tilde \j}|} e^{i (-\lambda_i + \tilde{\lambda}_{\tilde \j})}
\big{)}, \nonumber
\\
&f_{vec}(x, e^{i \lambda}, e^{i \tilde{\lambda}}) = -\sum_{i\neq
j}^N \big{(} e^{i (\lambda_i - \lambda_j)}x^{|n_i - n_j|}\big{)}-\sum_{{\tilde \i}\neq {\tilde \j}}^{\tilde N }
\big{(}  e^{i
(\tilde{\lambda}_{\tilde \i} - \tilde{\lambda}_{\tilde \j})} x^{|\tilde{n}_{\tilde \i} -
\tilde{n}_{\tilde \j}|} \big{)}, \nonumber
\\
&\epsilon_0 = \sum_{i,{\tilde \j}}  |n_i -\tilde{n}_{\tilde \j}| - \frac12
\sum_{i,j=1}^N |n_i - n_j| - \frac{1}2 \sum_{{\tilde \i},{\tilde \j} =1}^{\tilde N} |\tilde{n}_{\tilde \i}
-\tilde{n}_{\tilde \j}|.
\end{align}
The  index for the ABJ(M) model with $U(N)_k\times U({\tilde N})_{-k}$   is given by
\begin{align}
I_{ABJM}(x) = \sum_{\{n\},\{\tilde{n}\}} \frac{1}{(sym)}\int \frac{d^N
\lambda}{(2\pi)^N} \frac{d^{\tilde N} \tilde{\lambda}}{(2\pi)^{\tilde N}}\;
x^{\epsilon_0} \exp [i S_0 ] \exp \big{[} \sum \frac{1}{p} f_{tot}
(x^p, e^{i p \lambda_i },e^{i p \tilde{\lambda}_{\tilde \i}} )  \big{]}.
\label{abjmindex}
\end{align}
Here $(sym)$ is the symmetric factor, that is, the order of Weyl-group for
unbroken gauge group in the presence of monopole.
For example, $U(2)\times U(2)$ is broken to $U(2)\times U(1)$ due to the monopole $(n_1,n_2)=(1,1),
(\tilde{n}_1, \tilde{n}_2) = (1,2)$. In this case, the remaining Weyl-group is $S_2 \times S_1$ and
the symmetry factor
is $|S_2 \times S_1 | = 2! \times 1! = 2$.

The above superconformal index for small $N$ and $k=1$
can be calculated by using Mathematica at any order in $x$, and becomes
\begin{eqnarray}
&& I_{1_1\times 1_{-1}}(x) =  1 + 4 x^{1/2} + 10 x + 16 x^{3/2} + 19 x^2 + 20 x^{5/2} + 26
x^3 + 40 x^{7/2}   + 49 x^4 \nn \\ && \ \ \ \
+ 40 x^{9/2}
   + 26 x^5+  40 x^{11/2}    +84 x^6+100x^{13/2}+52 x^7 + 8 x^{15/2}+64 x^8+ 172 x^{17/2} \nn \\
   && \ \ \ \ + 150 x^9-16 x^{19/2}-61 x^{10}+172 x^{21/2}+376 x^{11}+152 x^{23/2} - 235 x^{12}+\cdots \label{u1ym} \\
&&
I_{2_1\times 2_{-1}}(x) = 1+4 x^{1/2}+20 x +56 x^{3/2}+139 x^2+260 x^{5/2} + 436 x^3 + 640 x^{7/2} + 954 x^4 + 1420 x^{9/2}\nn \\
&& \ \ \ \   +2076 x^5 + 2720 x^{11/2}+ 3234x^6 +
3780 x^{13/2}+ 5012 x^7+ 7048 x^{15/2}+ 8969 x^8+ \cdots  \label{u2ym}\\
&&
I_{3_1\times 3_{-1}}(x)=  1 + 4x^{1/2} + 20x + 76x^{3/2} + 239x^2 + 644x^{5/2} + 1512x^3 + 3100x^{7/2} + 5734x^4 \nn  \\
&&
\ \ \ \ + 9856x^{9/2} + 16182 x^5 + \cdots  \label{u3ym}\\
&&
   I_{4_1\times 4_{-1}}(x) =   1+4 x^{1/2}+20 x +76 x^{3/2}+274 x^2+844 x^{5/2} +2392x^3 + 6040 x^{7/2} \nn \\
&& \ \ \ \ + 13973 x^4 + 29456x^{9/2} +57756 x^5 +  \label{u4ym}\cdots
\end{eqnarray}

\subsection{the Index of the $U(N)$ Super Yang-Mills Theory}

Here we work out the index for the mirror dual of ${\cal N}=8$ $U(N)$ Yang-Mills theory. The
coincidence of the index convincingly reconfirms the equivalence between this ABJM  superconformal theory and
the IR limit of the ${\cal N}=8$ $U(N)$ Yang-Mills theory. The equivalence between them has been demonstrated
analytically by comparing the partition functions of two theories~\cite{Kapustin:2010xq}. However, the index
computation exhibits far more
information of the superconformal theories and we will find
quite an agreement.

For the  superconformal field theory of a  low energy super
Yang-Mills theory, we propose here the index formula which is the
naive generalization of the one (\ref{index structure}) given above.
At the IR limit, the inverse YM coupling $1/g_{YM}^2$ goes to zero
and so the contribution from Yang-Mills action vanishes, or  $S_0
=0$.  As commented at the subsection \ref{generalindex}, this could
be justified since the kinetic term of Yang-Mills is irrelevant in
the IR limit. The strategy works for many cases for the computation
of the partition function of the Yang-Mills theories with $\CN=4$
supersymmetry. But in some of the cases like $\CN=8$ super
Yang-Mills theories, naive localization leads to nonsensical
results, and so it needs a further consideration.   In the index
computation, we again consider the mirror dual,  the $\CN=4$ $U(N)$
Yang-Mills theory with one adjoint and one fundamental
hyper-multiplet.  This is the theory on $N$ D2 brane with a single
$D6$ brane. In the strong coupling limit the $D6$ brane near the
origin is lifted to a smooth geometry in M theory so we are left
with $N$ M2 branes. Thus, the relevant quantities for this $\CN=4$
theory is
\begin{eqnarray}
&& S_0 =0, \nonumber
\\
&& f_{hyper}(x, e^{i \lambda}) = \sum_{i=1}^N   \frac{x^{1/2}}{1+x}
x^{|n_i|}\big{(} e^{i \lambda_i } +   e^{-i\lambda_i} \big{)} +
\sum_{i,j=1}^N \frac{x^{1/2}}{1+x} x^{|n_i- n_j|} \big{(}e^{i
(\lambda_i - \lambda_j)} +e^{-i (\lambda_i - \lambda_j)}\big{)} ,
\nonumber
\\
&& f_{vec}(x, e^{i \lambda}) = -\sum_{i\neq j}^N  e^{i (\lambda_i -
\lambda_j)}x^{|n_i - n_j|}, \nonumber
\\
&& \epsilon_0 = \frac{1}2 \sum_{i=1}^N |n_i|.
\label{unindex0}
\end{eqnarray}
The Casimir energy due to the vector multiplet and the adjoint hyper-multiplet cancel each other.
The two hypermultiplets have the canonical conformal dimension 1/2. However, the adjoint chiral multiplet in the
$\CN=4$ vector multiplet has the conformal dimension 1 at the IR limit and so does not contribute to the index.
The full index for the IR conformal limit of the  $\CN=8$  $U(N)$ super Yang-Mills theory is then given
by
\be
I_{U(N)\ SYM}(x) = \sum_{\{n\} } \frac{1}{(sym)}\int \frac{d^N
\lambda}{(2\pi)^N}  \;
x^{\epsilon_0}   \exp \Big[ \sum_{p=1}^\infty \frac{1}{p} f_{tot}
(x^p, e^{i p \lambda_i }  )  \Big].
\label{unindex}
\ee
where the symmetric factor is again the order of the Weyl group of the
unbroken gauge group in the presence of magnetic charge.

We are claiming that the index (\ref{abjmindex}) of the ABJM model
for $U(N)_1\times U(N)_{-1}$ is identical to that (\ref{unindex}) of
the IR SCFT of the  3d $U(N)$ super Yang-Mills theory. We checked
extensively  the $x$ expansion of the index for  various small
values of $N$ and find a perfect agreement between them. This
agreement reaffirms the duality between the IR limit of $\CN=8$
super Yang-Mills theory with $U(N)$ gauge group, which is
regularized by a single fundamental hyper, and the ABJM model with
$U(N)_1\times U(N)_{-1}$. But it remains as a challenge to show this
equivalence in the exact analytic level.

One interesting case is $\CN=8$ $U(1)$ Yang-Mills theory. In the IR
limit this  describes   a single $M2$, which can be described by
(supersymmetric) free theory which consists of two hypermultiplets. The 4 complex scalars
in the theory correspond to  $\mathbb{C}^4$ where a single M2 branes
is probing. Thus we expect
that the index of $\CN=8$ $U(1)$ super Yang-Mills theory or
$U(1)_1\times U(1)_{-1}$ ABJM theory is the same as the free theory
with two hypermultiplets.  From index calculation, this can be
checked.  The superconformal index for the free theory can be
easily written as
\begin{eqnarray}
I_{free}(x ) = \exp [\sum_{n=1}^\infty \frac{1}{p} f_{free}(x^p)],
\quad f_{free} (x) :=   \frac{4x^{1/2}}{1+x}.
\end{eqnarray}
One can check that it is the same as the superconformal index for
$U(1)$ SYM and $U(1)_1\times U(1)_{-1}$ ABJM by the expansion. All the above computation can be done by
turning on the chemical potential. Again
the index matches.

To see this equivalence explicitly, we note that the  index for the ABJM model with $U(1)_1\times U(1)_{-1}$ is
\be I_{U(1)\ ABJM}= \sum_{n,\tilde n \in \IZ} x^{|n-\tilde n|}\int\frac{d\l}{2\pi}\frac{d\tilde \l}{2\pi}
e^{i(n\l-\tilde n\tilde\l)}\exp[\sum_{p=1}^\infty \frac1p f_{tot}(x^p,e^{ip\l},e^{ip\tilde \l}) ]
\ee
where
\be f_{tot}= \frac{2 x^{1/2}}{1+x} x^{|n-\tilde n|}(e^{i(\l-\tilde\l)} +
e^{-i(\l-\tilde\l)}) \ee
Now we put $r=\l-\bar\l, s=(\l+\tilde \l)/2$ with $\int d\l d\tilde \l /(2\pi)^2 = \int dr ds /(2\pi)^2$ with
the range for $\l,\tilde \l, r,s $ being $[-\pi,\pi]$. One can integrate over $s$ to get $n=\tilde n$ and
sum over $n$ of $e^{inr}$ leads to $\d(r)$ and so the $I_{U(1)ABJM}$ becomes the free theory index $I_{free}$.
So far, we only know the match in the series expansion  the explicit equivalence for the
index for the super Yang-Mills theory for $U(1)$ which is
\be I_{U(1)SYM}= \sum_{n\in \IZ} x^{|n|/2} \int_{-\pi}^{+\pi} \frac{d\l}{2\pi}
\exp\big(\sum_{p=1}^\infty \frac{1}{p} f_{tot}(x^p, e^{ip\l})\Big)
\ee
where
\be f_{tot} = \frac{x^{1/2}}{1+x} x^{|n|}(e^{i\l}+ e^{-i\l}) + \frac{2x^{1/2}}{1+x}
\ee

\subsection{Large N limit and the twisted sector}

One of the serious drawback of index computation is that we cannot
directly work out the index of ${\cal N}=8$ super-Yang-Mills theory
with gauge group $G$.  At the technical level, this is obvious from
the index formula in eq.~\eqref{unindex0}. The Casimir energy for
any monopole operators vanishes because contributions from an
adjoint hyper and from ${\cal N}=2$ vector multiplet cancel each
other. In addition, there're no CS terms in the SYM theory and thus
monopole operators are gauge-invariant by themselves so they  need
not be combined with charged matters.  The index gets divergent as
we take the sum over all these energy-zero monopole operators. Thus
we cannot compute the index for $N$ D2s directly.

However one can compute the index for the field theory corresponding
to $N$ D2s with $m$ D6s for arbitrary nonzero $m$. The $m$ D6s introduce $m$ fundamental
hypermultiplets to ${\cal N}=8$  super-Yang-Mills theory and break supersymmetries  to ${\cal N}=4$.
Fundamental hypermultiplets give positive Casimir energy to monopole operators
and  make the superconformal indices of ${\cal N}=4$ theories to be finite.
The vacuum moduli space
for such theories are composed of the geometric branch from vector and adjoint hyper multiplets and
the Higgs branch from fundamental  hyper multiplets. We are interested in here the geometric branch.
The vector multiplet for $N=1$ has the charge $m$ Taub-Nut space as moduli space.  The explicit metric can be written as
\begin{align}
d s^2_{\textrm{Taub-Nut}} &= H  d\vec{r}\cdot d \vec{r} + H^{-1} ( d\tau+ \vec{\omega}\cdot d\vec{r})^2 \;,
\textrm{where} \nonumber
\\
H &= \frac{1}{g^2} + \frac{m}{|\vec{r}|}, \quad \vec{\nabla} H = \vec{\nabla} \times \vec{\omega}.
\label{Taub-Nut metric}
\end{align}
Classical moduli space $\mathbb{R}^3\times S^1$ is 1-loop corrected by integrating out $m$ fundamental hypermultiplets. Here $\tau$ denote the dual photon coordinate with peridicity $4\pi$ which corresponds to $2 \phi_8^a$ in \eqref{dual photon}. The second term in the harmonic function $H$ is due to the 1-loop effect \cite{Seiberg:1996nz, Intriligator:1996ex}.
 In the IR limit, where gauge coupling $g$ goes to infinity, the classical part in
the harmonic function disappears and  the geometry become  $\mathbb{C}^2/ {\mathbb Z}_m$. Combining the moduli space
from adjoint Higgs, total geometric branch become $\mathbb{C}^2 \times (\mathbb{C}^2/\mathbb{Z}_m )$ for $N=1$.
The additional Higgs branch other than
geometrical branch gives index from twisted sector, which is discussed below.

   The geometric moduli space for the IR limit with general $N$ would be the symmetric product of
   $\IC^2\times (\IC^2/\IZ_m)$.
Its gravitational dual in the large N limit
is given by $AdS_4 \times S^7/{\mathbb Z}_m$. Thus we are in the
situation where we know how to compute the field theory index
associated with the orbifolded daughter theories in the field theory and gravity sides
while we do not know the field theory index for the†
parent theory before the orbifolding. Our strategy is by working out
the various daughter theories and infer indirectly on the parent
theory.

Concretely, the gravity index on $AdS_4 \times S^7/{\mathbb Z}_m$ is
given by the ${\mathbb Z}_m$ invariant projection of the index on
$AdS_4 \times S^7$ and the twisted sector contribution
\cite{Sangmin:2008}. Thus if we subtract the twisted sector
contribution from the field theory index dual  to $AdS_4 \times
S^7/{\mathbb Z}_m$, we come to know the $\mathbb{Z}_m$ invariant
contribution of the field theory associated with $N$ D2s.  By
working out the index for arbitrary $m$ we can deduce the field
theory index associated with $N$ D2s. The twisted sector
contribution comes from D6-D6 states and it is known how  to compute
in the gravity side \cite{Imamura:2009hc}, \cite{Imamura:2010sa} .
Thus by working on this scheme, one can indirectly work out the
index on $N$ D2s in the large N limit, which is index of ${\cal
N}=8$ $U(N)$ Yang-Mills theory. We will adopt the similar strategy
later for other gauge groups.  In this way, we can establish the
equivalence between the IR limit of ${\cal N}=8$ $U(N)$ Yang-Mills
theory and ABJM theory in Large $N$ limit. By considering the
Higgsing pattern one can see that the equivalence should hold for
finite $N$ as well.

It is convenient to turn on the chemical potential $y$ for the $U(1)_{diag}$ monopole
charge $h$ and define the index as
\begin{equation}
I(x,y)= \Tr (-1)^F x^{\epsilon+j_3} y^h, \quad h = m \sum_{i=1}^N n_i \;.
\end{equation}
Let us consider ${\cal N}=4$ theory associated with $N$ D2s and $m$ D6s. This is
described by adding $m$ fundamental hypermultiplets to $\mathcal{N}=8$ $U(N)$ Yang-Mills theory.
The  geometric branch  of moduli space for a single D2 is described by
$\mathbb{C}^2 \times \mathbb{C}^2/ \mathbb{Z}_m$. One can consider the global rotation
\begin{equation}
(z_1,z_2,z_3,z_4) \rightarrow (z_1,z_2, e^{i\theta}z_3
e^{-i\theta}z_4)
\end{equation}
where this action corresponds to translation in $\tau$-direction in \eqref{Taub-Nut metric}.
Recalling that the $\tau$ is the dual photon direction,  quantum number
under this action can be  identified with the monopole charge $h$.
In M-theory picture, D6 branes are mapped to KK monopole ($\mathbb{R}^{1,6} \times $ Taub-Nut)
and the 11-th dimensional circle corresponds to the $\tau$ coordinate in the Taub-Nut space.

The detailed field theoretic computation of the large N limit is relegated to the
appendix \ref{appB1}. Here we summarizes the results of the
computation. Let $I_{U(\infty):m}(x,y)$ be the large N index for $U(N)$
Yang-Mills theory with $m$ fundamental hypermultiplets. The salient
feature of the index is that it is factorized into
\begin{equation}
I_{U(\infty):m}(x,y)=I_{U(\infty):m}^{(0)}(x)I_{U(\infty):m}^{(+)}(x,y)I_{U(\infty):m}^{(-)}(x,y),  \label{decomp}
\end{equation}
where $0,+,-$ denotes the zero, positive, negative monopole charge sectors,
respectively. From the gravitational side this is mapped to graviton
index with zero, positive, negative KK momentum sectors
respectively. From the fact that this gives graviton index, it
should be written as Plethystical form
\begin{equation}
I_{U(\infty):m}(x,y)= \exp \sum_{n=1}^{ \infty} \frac{1}{n} I_{U(\infty):m:sp}(x^n, y^n),
\end{equation}
where $I_{m;sp}$ denotes the single particle index. From the
decomposition of eq. (\ref{decomp}), the single particle index can
be written as
\begin{eqnarray}
I_{U(\infty):m:sp} &= &I^{(0)}_{U(\infty):m:sp}+ I^{(+)}_{U(\infty):m:sp} + I^{(-)}_{U(\infty):m:sp} \;,  \nonumber
\\
& =& \sum_{n=-\infty}^\infty  y^{mn}I^{(mn)}_{U(\infty):m:sp} (x) \; .
\end{eqnarray}
Again  $0,+,-$ denotes the zero, positive, negative monopole sector,
respectively. For the zero momentum sector, one can explicitly work
out the single particle index in field theory and is given in eq.~\eqref{I^(0)_{U(inf)}}:
\begin{align}
 I^{(0)}_{U(\infty):m:sp} (x)
 &=\frac{ 2 (x^{1/2} +  x -  x^{5/2}) }{ 1- x^{1/2} - x^2 + x^{5/2} } + (m^2 -1) \frac{x}{(1+x) (1- x^{1/2})^2  }  \ .
 \label{index:m=0:sp}
 \end{align}
%
Especially for $m=1$ case, this
coincides with the single particle gravity index with zero momentum
on $AdS_4\times S^7$   given in eq.~\eqref{I^{(n)}_{S^7;sp}}.

From the gravitational dual perspective, the
field theory index can be decomposed into
\begin{equation}
I_{U(\infty):m:sp}(x,y)=I_{S^7/\mathbb{Z}_m:sp}(x,y)+I_{U(\infty):m:twisted:sp}(x,y) \ . \label{U(N) bulk+twisted}
\end{equation}
In the gravity side the bulk index $I_{S^7/\mathbb{Z}_m:sp}(x,y)$
comes from the single graviton index on $AdS_4 \times S^7$ by
keeping the invariant states under the $\mathbb{Z}_m$ orbifolding,
\begin{equation}
I_{S^7/\mathbb{Z}_m:sp}(x,y)=I^{(0)}_{S^7:sp} (x)+\sum_{n\neq 0}y^{mn} I^{(mn)}_{S^7:sp}(x) \label{index:bulk}
\end{equation}
where $I_{S^7:sp}^{(n)}$ denotes the gravity index on $AdS_4 \times
S^7$ with $n$ units of KK momentum along 11-th circle. The twisted sector comes from
the fixed locus under the $\mathbb{Z}_m$ action, which is $AdS_4
\times S^3$. D6 branes supported on the fixed locus provide this
contribution. For a single D6 brane, the worldvolume theory is
seven-dimensional maximal supersymmetric theory on $AdS_4 \times
S^3$, which consists of a single vector multiplet. Since $AdS_4
\times S^3$ is embedded into the eleven-dimensional $AdS_4 \times
S^7$, so that we should use the M-theory picture. However since the
world volume theory does not probe the 11-th circle, usual D6 brane
picture does make sense since the spectrum should be independent of
the radius of the 11-th circle. The spectrum is worked out in
\cite{Imamura:2009hc, Imamura:2010sa}. For a single
D6, 
the index of the 7-d world-volume theory is given as
 \be I_{sp}^{U(1); AdS_4 \times S^3} (x)  =
\frac{x}{1-x^2 } ( 1 + 2 \frac{ x^{1/2} }{ 1- x^{1/2}}) =
\frac{x}{(1+x)(1-x^{1/2})^2 }.  \ee
For $m$ D6 branes, we have $SU(m)$
gauge groups and we have
\begin{equation}
I_{sp}^{SU(m); AdS_4 \times S^3} (x)  = (m^2 -1)
\frac{x}{(1+x)(1-x^{1/2})^2 } \label{D6 index}
 \end{equation}
which exactly matches with the second term in eq.~\eqref{index:m=0:sp}.
%
%
By assuming the equality between field theory large N index for $m=1$  with the gravity index on
$AdS_4 \times S^7$, which was extensively checked,
\begin{equation}
I_{U(\infty):m=1:sp}(x,y)=I_{S^7:sp}(x,y). \label{assumption U(N)}
\end{equation}
one can indeed show that twisted sector index defined in \eqref{U(N) bulk+twisted}
are exactly same with the index fromt $m$ D6 branes \eqref{D6 index}.
From eq.~\eqref{I^{(n)}_{S^7;sp}}, one can see that the
graviton index satisfies $I_{S^7 : sp}^{(mn)}=x^{ \frac{(m-1)|n|}{2}
} I_{S^7: sp}^{(n)} (x) $.  On the other hand, from
eq.\eqref{nonzero-monopole} the large N index satisfies
$I_{U(\infty):m:sp}^{(mn)} = x^{\frac{(m-1)|n|}2
}I^{(n)}_{U(\infty):m=1:sp}$ for $n\neq 0$ . Thus, under the the assumption \eqref{assumption
U(N)} one can see that $I_{S^7;sp}^{(mn)} = I_{U(\infty):m:sp}^{(mn)}$ for
$n\neq 0$. This imply that twisted sector index in
\eqref{U(N) bulk+twisted}  is
\begin{eqnarray}
I_{U(\infty):m:twsted:sp} (x,y) &=& \sum_{n=-\infty}^\infty (I^{(mn)}_{U(\infty):m:sp} - I^{(mn)}_{S^7}) y^{mn} = I^{(0)}_{U(\infty):m:sp} (x) - I^{(0)}_{S^7} (x) \;, \nonumber
\\
&=& (m^2 -1) \frac{x}{(1+x)(1-x^{1/2})^2 } .
\end{eqnarray}
In the last line  we use explicit expression of $y^0$ part of large N field theory index \eqref{index:m=0:sp} and graviton index  on $AdS_4\times S^7$ \eqref{I^{(n)}_{S^7;sp}}. This exactly match with index form D6s \eqref{D6 index} as claimed. One peculiar   feature is that in the field theory index the
twisted sector contribution comes only from the zero-monopole
sector. This is sensible since the monopole charge is identified
with KK momentum along the 11-th circle and the twisted sector on
$AdS_4 \times S^3$ cannot probe the 11-th circle by construction.
 One can explicitly construct local operators in SYM theory corresponding to twisted sector index.
 Listing some lowest order example,
\begin{align}
&A_I B_J  \textrm{ modulo } F_\Phi = B_I A_I\sim 0 \quad : \quad (m^2-1)x, \nonumber
\\
&A_I \Phi_1 B_J, \;  A_I\Phi_2 B_J \textrm{ modulo } F_\Phi \sim 0 \quad : \quad 2(m^2-1) x^{3/2}, \ldots
\end{align}
Here $F_\Phi$ denote the $F$-term condition for adjoint chiral
superfield $\Phi$ in $\mathcal{N}=4$ vector multiplet. $\Phi_1$,
$\Phi_2$ denote chiral superfields in adjoint $\mathcal{N}=4$
hyper-multiplet and $A_I$, $B_I$ are chiral superfields in  $m$
fundamental hyper-multiplet.    Thus overall arguments lead that
field theory index defined on $N$ D2s coincide with the gravity
index on $AdS^4 \times S^7$ in the large N limit, which is the same
as the index of ABJM theory with Chern-Simons level 1. It would be
desirable to work out the index on $N$ D2s directly.

\subsection{Higgsing of ABJM and super Yang-Mills}

In the previous subsection we show that the index of large $N$ limit
of $U(N)$ Yang-Mills theory is the same as that ABJM theory with
$k=1$. It's important to extend this equality to finite $N$. For
$U(N)$ case, we resort to the mirror dual of ${\cal N}=8$, $U(N)$
Yang-Mills whose index is the same as that of ABJM for any finite
$N$. This method is not available for other gauge groups. Thus we
look for the Higgsing pattern for Yang-Mills theory and ABJM and
induce the equivalence of the two theories.  Starting from $U(N)
\times U(N)$ ABJM one can consider the Higgsing to $U(1)^{N}\times
U(1)^N$ gauge theory. This corresponds to separating all of M2
branes. If the interdistance between any of M2 branes are very
large, we expect that we obtain product of free theorie in the IR
limit.

Let us turn on the vev of one complex scalar $Z_4$ of $U(N) \times
U(N)$ ABJM
\begin{equation}
Z_4=\bar{Z}^4=\left( \begin{array}{ccc}
                R_1 & 0 & 0\cdots \\
               0 & R_2 & 0 \cdots \\
               0 & 0 & R_3 \cdots    \end{array}  \right)
\end{equation}
According to \cite{HosomichiYi} et al, all of the offdiagonal
components have the mass $\frac{2\pi}{k}(R_i^2-R_j^2)$.  We are
taking the limit $R_i, R_j \rightarrow \infty$ with $R_i\neq R_j$.
We expect that all of these massive modes are decoupled in such
limit. Thus we are left with abelian ABJM and it is sufficient to
see what happens to $U(1) \times U(1)$ ABJM with turning on
bifundamental vev. $U(1)\times U(1)$ ABJM is given by

\begin{equation}
L=\frac{k}{4\pi} (A_{\mu}\partial_{\nu}
A_{\rho}-\tilde{A}_{\mu}\partial_{\nu}
\tilde{A}_{\rho})-D_{\mu}\bar{Z}^{\alpha}D^{\mu}Z_{\alpha}-i\bar{\Psi}_{\alpha}D_{\mu}\Psi^{\alpha}
\end{equation}
with
\begin{equation}
D_{\mu}Z_{\alpha}=\partial_{\mu}Z_{\alpha}-i(A_{\mu}-\tilde{A}_{\mu})Z_{\alpha}
\end{equation}
With $Z_{\alpha}=R\delta^{4}_{\alpha}+Y_{\alpha}$, we integrate out
$A_{\mu}-\tilde{A}_{\mu}$  to obtain
\begin{eqnarray}
&
&L=-D_{\mu}\bar{Z}^{\alpha}D^{\mu}Z_{\alpha}-i\bar{\Psi}_{\alpha}D_{\mu}\Psi^{\alpha}
\nonumber \\
& &
+\frac{(\frac{k}{4\pi}\epsilon^{\mu\nu\rho}\partial_{\nu}(A_{\rho}+\tilde{A}_{\rho})
+iR(\partial_{\mu}(\bar{Y}^4-Y_4)+\bar{\Psi}_{\a}\gamma_{\mu}\Psi^{\alpha}
+i\partial_{\mu}\bar{Y}^{\alpha}Y_{\alpha}-i\bar{Y}^{\a}\partial_{\mu}Y_{\a})^2}{4(R^2+R(Y_4+\bar{Y}_4)+Y_{\a}\bar{Y}^{\a})}
\end{eqnarray}
Replace $A_{\rho}+\tilde{A}_{\rho}$ by
$R(A_{\rho}+\tilde{A}_{\rho})$ to obtain the standard kinetic term
for the gauge field, we obtain
\begin{equation}
L=-\frac{k^2}{32\pi^2}(F+\tilde{F})^2-\sum_{\alpha=1}^3\partial_{\mu}\bar{Z}^{\alpha}\partial^{\mu}Z_{\alpha}
-\frac{1}{2}(\partial
\frac{\bar{Y}^4+Y_4}{\sqrt{2}})^2-i\bar{\Psi}_{\alpha}\partial_{\mu}\Psi^{\alpha}
+O(\frac{1}{R})
\end{equation}
which is ${\cal N}=8 U(1)$ Yang-Mills theory. Thus in the limit
$R_i, R_j \rightarrow \infty$ with $R_i \neq R_j$ the low energy
theory of $U(N)\times U(N)$ ABJM with level $k$ is given by $N$
copies of ${\cal N}=8 \, U(1)$ Yang-Mills theory.

 Now consider $\CN=8 \, U(N)$ super Yang-Mills theory and
$U(N)_1\times U(N)_{-1}$ ABJM theory. They have the same moduli
space and the same large N limit.  Also Higgsing pattern is
consistent.  From the Higgsing from  $U(N)$ Yang-Mills, we can
obtain $U(1)^N$ Yang-Mills.  In the ABJM side, we obtain the same
$U(1)^N$ Yang-Mills theory after the Higgsing. After the Higgsing,
they have the same index for a trivial reason. In BCD case, this
kind of argument is effective in telling which Yang-Mills theory
should be mapped to which ABJ type theory.

\section{ BCD super Yang-Mills   and $k=2$ ABJ(M) Models }

We are proposing the dualities between the IR limit of $\CN=8$ super
Yang-Mills theories and the ABJ(M) models. In this section we are
interested in the following duality
\begin{equation} \left. \begin{array}{rcclr}
 O(2N)    & {\rm SYM} &  \Longrightarrow  &  U(N)_{2} \times U(N)_{-2}  & {\rm ABJM} \\
 SO(2N+1)  & {\rm SYM} &  \Longrightarrow  &  U(N)_{2} \times U(N+1)_{-2}  & {\rm ABJ} \\
 Sp(2N)  &  {\rm SYM}  & \Longrightarrow &  U(N)_{ 2} \times U(N)_{-2}  & {\rm ABJM} \\
  &    &  \Longrightarrow  &  U(N)_{2} \times U(N+1)_{-2}  & {\rm ABJ}  \\
\end{array}\right. . \end{equation}
In this section, we  want to test the proposal in Sec.2 for the
correspondence between   the infrared limit of the ${\cal N}=8$
supersymmetric Yang-Mills theories of BCD type gauge group and the
ABJ(M) models, which is summarized in Table II.  The main tool is to
compare  the indices of ABJ(M) models and those of super Yang-Mills.
Unlike the previous section for $U(N)$ super Yang-Mills theory,
there is no clear regularization process for the BCD case  and we
improvise the index for small $N$, and use the large $N$ limit of
the field theory and gravity calculation to infer the index.  For
$N=2$ the duality is proposed in \cite{Lambert:2010ji}.

Let us first calculate  the indices (\ref{abjmindex}) of the ABJ(M) model of
$ U(N)_k\otimes U(\tilde N)_{-k}$ for $k=2$  for small $N$:
\begin{eqnarray}
&&I_{ABJM}[ 1_2, 1_{-2}   ] = 1+ 10 x + 19 x^2 + 26 x^3 + 49
x^4 + 26 x^5 + 84 x^6+ 52 y^7 +  \cdots \label{ao2index}  \\
&&I_{ABJM}[1_{2},2_{-2}]  =1+10 x+20 x^2+20 x^3+65 x^4+10 x^5+55 x^6+190 x^7+\dots ,   \label{aso3index}\\
&&I_{ABJM}[2_{2}, 2_{-2}] =1+10 x+75 x^2+220 x^3+475 x^4+1060 x^5+1665 x^6 + \dots. \label{ao4index} \\
&&I_{ABJM}[2_{2}, 3_{-2}] =1+10 x+75 x^2+230 x^3+449x^4 + 1026 x^5 +1990 x^6   \dots.\\
&&I_{ABJM}[3_{2}, 3_{-2}] =1+10 x+75 x^2+450 x^3+1595 x^4+4230 x^5+ \ldots. \label{ao6index}\\
&&I_{ABJM}[3_{2}, 4_{-2}] =1+10 x+75 x^2+450 x^3+1650 x^4+4240 x^5+ \dots.
\end{eqnarray}

\subsection{Index of $\CN=8$ BCD Yang-Mills Theory for small rank gauge groups}

We calculate the indices for $O(2N), SO(2N+1), Sp(2N)$ for small
$N$. Before we consider these cases, let us consider the index
(\ref{unindex}) for the super Yang-Mills theory for $SU(N)$ gauge
group. Note that
\be U(N) = \frac{SU(N)\otimes U(1)}{{\mathbb Z}_N} \ee
Thus we propose  the index for $SU(N)/{{\mathbb Z}_N}$ $\CN=8$ SYM theory to be
\be I_{SU(N)/{\mathbb Z}_N} = \frac{I_{U(N)}}{I_{U(1)}} \ee
As there are only adjoint matter fields in the $\CN$ theory, the above index is really the index of $SU(N)$ $\CN=8$
SYM theory.
For example
\be   I_{SU(2)/{\mathbb Z}_2} &=& 1+ 10 x + 20 x^2+20 x^3+65 x^4+10 x^5+55 x^6 + 190 x^7+\cdots    \label{su2index} \\
  I_{SU(3)/{\mathbb Z}_3}&=&  1+ 10 x + 20x^{3/2} +40 x^2 + 104 x^{5/2}+ 160 x^3 + 361 x^4 + 516 x^{9/2} + \cdots   \\
  I_{SU(4)/{\mathbb Z}_4}&=& 1+  10 x + 20 x^{3/2}+75 x^2+164 x^{5/2}+450 x^3+ 780 x^{7/2} \nn \\ & & + 1595 x^4
  +2500 x^{9/2}+4230 x^5+\cdots \label{su4}
\ee
and so on.

Let us now consider the simplest case, the $\CN=8$ super Yang-Mills theory with
$O(2)=SO(2)\otimes {\mathbb Z}_2= U(1)\otimes \IZ_2$, where ${\mathbb Z}_2$ acts on
a complex scalar by it complex conjugation  and the sign change for an adjoint scalar.
The fractional power in the index for $U(1)$ gauge group denotes the odd number of fields.
Thus we can regard the $O(2)$ index to be that of $U(1)$ index without fractional power:
\be  I_{O(2)}(x) &=& 1   + 10 x   + 19 x^2   + 26 x^3     + 49 x^4
   + 26 x^5        +84 x^6 \nn \\
   & & + 52 x^7  +64 x^8  +150 x^9 -61 x^{10}+ 376 x^{11}  - 235 x^{12}+\cdots
   \ee
This matches the index (\ref{ao2index}) of ABJM with $U(1)_2\times U(1)_{-2}$ exactly.

The index for $O(4)$ needs a bit more consideration. First of all $SO(4)= SU(2)\times SU(2)/\IZ_2 $ for
the $4d$ real vector representation. For the adjoint representation $SO(4)=SO(3)\times SO(3)$.
Thus $O(4)=SO(3)\times SO(3)\times \IZ_2$. The vacuum moduli space of the IR dynamics of $O(4)$ super
Yang-Mills theory is
\be \CM_{O(4)} = (\IC^4/\IZ_2)^2/\IZ_2= (\CM_{SO(3)})^2/\IZ_2 \ee
Thus, we expect that $O(4)$ index to be that of two particle index of $SO(3)$:
\be I_{O(4)\ SYM}= \frac12 \Big\{ I_{SO(3)\ SYM}(x^2) + \Big[ I_{SO(3)\ SYM} (x)\Big]^2\Big\} \ee
which matches exactly that (\ref{ao4index}) of $U(2)_2\times U(2)_{-2}$ ABJM model.

The index for the $O(6)=SO(6)\times \IZ_2$ super Yang-Mills theory can be obtained from that (\ref{su4})
for $SO(6)= SU(4)$ by dropping the fractional power similar to the $O(2)$ case:
\be I_{O(6)} = 1+  10 x +  75 x^2 +450 x^3    + 1595 x^4 +4230 x^5+\cdots \label{o6}
\ee
which matches the index (\ref{ao6index}) of ABJM model for $U(2)_2\times U(3)_{-2}$.

The index for the $SO(3)=Sp(2) $ case  is the index   for the
$SU(2)$ super Yang-Mills theory, which is $I_{U(2)}/I_{U(1)}$. The
direct calculation (\ref{su2index})  matches the index
(\ref{aso3index}) of ABJM model for $U(1)_2\times U(2)_{-2}$.  On
the other hand we know that for $SU(2)=Sp(2)$ there could be another
superconformal field theory where this theory flows. There could be
subtleties in taking the IR limit.  In the low energy limit, the
gauge group is reduced to $U(1)$ with seven scalars $\phi_i$ and
$\phi_8$, dual scalar to a photon. The moduli space with the finite
coupling is given by $\CM= \frac{{\mathbb R}^7\times S^1}{\mathbb{Z}_2}$ as a
special case of (\ref{modun2}). It has two singularities at
$\phi_i=\phi_8=0$ and $\phi_i=0, \phi_8=\pi$. In the infinite
coupling limit, the theory at two orbifold singularities has the
moduli space of ${\mathbb R}^8/\mathbb{Z}_2$. It is argued in \cite{Seiberg},
the theory at $\phi_i=\phi_8=0$ leads to interacting conformal
theory while the theory at $\phi_i=0, \phi_8=\pi$ is a free field
theory with a gauged $\mathbb{Z}_2$ symmetry. Note that $\phi_8$ corresponds
to the position of M2 brane in the 11-th circle and the different
value of $\phi_8$ implies the different $OM2$-plane in the M-theory
setting. Note that for the theory defined at $\phi_8=\pi$, the $\mathbb{Z}_2$
flips the sign of the scalar field $\phi_i, \phi_8$ so that only the
operators of the even $\phi_i, \phi_8$ will survive. Thus from the
free field theory index we have to remove the operators having
half-integer powers. The resulting index is the same as that of
$U(1)_2\times U(1)_{-2}$ ABJM theory. The other superconformal field
theory living at $\phi_i=\phi_8=0$ can be identifield with $U(1)_2
\times U(2)_{-2}$ ABJ theory.

To summarize, we explicitly checked the indices of the following cases:
\be \begin{array}{|r|l|l|}
\hline
    {\rm SYM}  &  {\rm Index} & {\rm ABJM} \\
    \hline
    O(2) & I_{U(1)}|_{no fraction} & U(1)_2\times U(1)_{-2} \\
    SO(3)& I_{SU(2)} & U(1)_2\times U(2)_{-2} \\
    O(4) &  ( I_{SU(2)}(x^2)+ [ I_{SU(2)}(x)]^2)  /2 & U(2)_2\times U(2)_{-2} \\
    O(6) & I_{SU(4)}|_{no fraction} &    U(3)_2\times U(3)_{-2} \\
    Sp(2) & & U(1)_2\times U(1)_{-2}  \\
     &  & U(1)_2 \times U(2)_{-2} \\
    \hline
    \end{array} \ee
It would be desirable to define the indices for these BCD
class for larger $N$. Adding one fundamental hyper-multiplet does
not seem to work unlike the $U(N)$ case. But this leads to the $Z_2$
orbifold theories we will consider in the next subsection.

\subsection{$\mathbb{Z}_2$ Orbifolded Theories}

Obviously the above computation can be done only for small ranks of
the gauge group.  Our roundabout way of the computation consists of
several steps. The first step is to work out $\mathbb{Z}_2$ orbifolded
theories of $\CN=8,  O(2N)/Sp(2N)$ theories and match to a suitable
Chern-Simons dual for any $N$. The second step is to work out $Z_m$
orbifolded theories of  $ \CN=8, BCD$ theories and show that these
theories are dual to the gravity theory on $AdS_4 \times S^7/\mathbb{Z}_2$ in
the large $N$ limit after taking account of the twisted sector
contribution to the index. From the first and the second step, one
can argue the equivalence between $\CN=8, O(2N)/Sp(2N)$ and ABJ(M)
theory with $k=2$ for any finite $N$. In this calculation, $Sp(2N)$
naturally matches to $U(N)_{2} \times U(N+1)_{-2}$. For $SO(2N+1)$
and another branch of $Sp(2N)$, we rest on the Higgsing pattern to
check the consistency of the proposed dualities.

Let us look for the theory with $2N$ D2-branes with $2m$ D6-branes
and carry out the orientifold projection. This gives
$O(2N)/SO(2N+1)/Sp(2N)$ gauge group for D2s. Consider the brane
system $2N$ D2s (012) and $2m$ D6s (012345). Lifting to M-theory,
D2s become M2-branes probing $\mathbb{C}^2/ \mathbb{Z}_{2m} \times
\mathbb{C}^2  $. Matter fields in the D2/D6 system are
\begin{align}
\begin{array}{ccc}
   & U(2N) & U(2m) \\
  \textrm{Hyper} & Adj &  \\
  \textrm{Hyper} &  & Adj \\
  \textrm{Hyper} & 2N & \overline{2m} \\
   & \overline{2N} & 2m  \\
\end{array}
\end{align}
$U(2N)$ is the gauge symmetry on D2-branes and $U(2m)$ is the gauge
symmetry on D6s, which is the global symmetry of the D2-brane
world-volume theory. After introducing O2$^{-}$ (012), the geometry
probed by M2-branes  becomes $(\mathbb{C}^2/\mathbb{Z}_{2m} \times
\mathbb{C}^2 )/\mathbb{Z}_2$. Matter fields in the system are
projected into
\begin{align}
\begin{array}{ccc}
   & O(2N) &  Sp(2m) \\
  \textrm{Hyper} & Adj &  \\
  \textrm{Hyper} &  & Adj \\
  \textrm{(real)Hyper} & 2N & 2m \\
\end{array}
\end{align}
Combining  each pair of $2m$ (real) hyper-multiplets, one can make
$m$ hypermultiplets. The world-volume theory of D2s is given by
$O(2N)$ $\mathcal{N}=4 $ SYM with hyper-multiplets, one in the
adjoint and $m$ in $2N$ vector representation. In the IR limit, the
theory becomes $N$ M2-branes' world-volume theory on
$(\mathbb{C}_1^2 /\mathbb{Z}_{2m} \times \mathbb{C}_2^2
)/\mathbb{Z}_2$. The action of the discrete quotient on
$\mathbb{C}_1^2 \times \mathbb{C}_2^2$ is given by
\begin{align}
&\alpha =  \exp (\frac{4 \pi i J_3}{2 m}) \otimes \mathbb{I} , \nonumber
\\
& \beta = \exp (\pi i J_2 ) \otimes (- \mathbb{I} ) =\exp (\pi i J_2 )  \otimes \exp (2 \pi i J^\prime_3). \label{alpha and beta}
\end{align}
Here $\alpha, \beta$ is the generators of
$\mathbb{Z}_{2m},\mathbb{Z}_2$ respectively. $\{J_i , J_i'\}$  are
generators of $SU(2), SU(2)^\prime$ which act on $\mathbb{C}^2_1$
and $\mathbb{C}^2_2$ respectively. Focusing only on the first
$\IC^2_1$ factor in $\IC^4$, $(\alpha,\beta)$ generate dihedral
group $D_{m}$ (4m elements) action on $\IC^2_1$.

On the other hand, related world-volume theory of M2s on
$(\mathbb{C}^2/\mathbb{Z}_{2m} \times \mathbb{C}^2)/\mathbb{Z}_2$
can be obtained from the Hannay-Witten setup in Type IIB theory with
$N$ D3 branes, $2m$ NS5-branes and an
$(1,2)$-brane\cite{Imamura:2008nn}. But in this case, the
$\mathbb{Z}_2$ action is different from the $O(2N)$ SYM. The
generator of the $\mathbb{Z}_2$, say $\tilde{\beta}$, act on $\IC^4$ as
\begin{align}
\tilde{\beta} = \exp (\frac{4 \pi i J_3}{4 m})\otimes \exp (2 \pi i J'_3).
\end{align}
In this case, $(\alpha, \tilde{\beta})$ form cyclic group
$\mathbb{Z}_{2\times 2m} = \mathbb{Z}_{4m}$ action
on the first $\mathbb{C}^2_1$.
$m=1$  is a special case, when $D_m =\mathbb{Z}_{4m}$ and group action generated by $(\alpha,\beta)$ are equivalent to action generated by $(\alpha,\tilde{\beta})$ up to some basis change in $\IC^4$ . Note that
2 NS5-branes with an (1,2)-brane gives rise to $U(N)_2\times
U(N)_0\times U(N)_{-2}$ ${\cal N}=4$ SCSM with bifundamental
matters $(N, \bar{N},1), (1, N, \bar{N}), (\bar{N},1,N)$. Thus we
suggest
\begin{align}
&\textrm{$\mathcal{N}=4$ $O(2N)$ SYM with one hyper in adjoint and
one in $2N$ vector } \nonumber
\\
&\Longrightarrow_{IR} \; \textrm{ $\mathcal{N}=4$ SCSM with gauge
group $U(N)_{2}\times U(N)_{-2}\times U(N)_0$ }.
\end{align}
 We confirm the duality by comparing superconformal
index. One subtlety in the calculation is that the SYM gauge group
is $O(2N)$ but not $SO(2N)$. After taking this subtlety
\cite{Scheon:2011},\cite{Hwang:2011}, the index of the SYM for $N=1$
as an example, is given by
\begin{align}
  I (x)   = 1 + 7 x + 4 x^{3/2} + 16 x^2 + 4 x^{5/2} + 21 x^3 + 8 x^{7/2} +
 40 x^4 + 28 x^{9/2} + 34 x^5 + \ldots
\end{align}
This matches the index of  SCSM with $U(1)_{2}\times
U(1)_{-2}\times U(1)_0$.

Instead of considering $O2^-$ one can consider $\tilde{O2}^+$. In
this case we have $Sp(2N) \times SO(2m)$ gauge group for $N$ D2 and
$m$ D6s. By the similar logic, one can see that one can compare
$Sp(2N)$ SYM to the Chern-Simons type theory. In this case the IR
limit of the $Sp(2N)$ SYM with one fundamental hyper-multiplet is
equivalent to the ${ \cal N}=4$ SCSM with gauge group $U(N)_2\times
U(N+1)_{-2}\times U(N)_0$:
 \begin{align}
&\textrm{$\mathcal{N}=4$ $Sp(2N)$ SYM with one hyper in adjoint and
one in $2N$ vector } \nonumber
\\
&\Longrightarrow_{IR} \; \textrm{ $\mathcal{N}=4$ SCSM with gauge
group $U(N)_{2}\times U(N+1)_{-2}\times U(N)_0$ }.  \label{z2claim}
\end{align}
 Note that in the index computation of $Sp(2N)$, we implicitly choose the value of $\theta=0$
 in (\ref{basicaction}). We do not know how to implement the other value in the index computation.
We also check (\ref{z2claim}) by the index. Again as an example, the
index of ${ \cal N}=4$ $Sp(2)$ SYM
 and $U(1)_2 \times U(2)_{-2} \times U(1)_0$
SCSM coincide with each other, given by
\begin{align}
I (x)
& = 1+ 5x + 8 x^{3/2} + 9 x^2 + 12 x^{5/2} + 16 x^3 + 4 x^{7/2} + 29 x^4 + 56 x^{9/2} + O(x^5) \ .
\nn
\end{align}
On the other hand, we attempt to find similar theory for $SO(2N+1)$
theory, but we do not succeed. This would be an interesting problem
to find such theory.

\subsection{Large $N$ limit and  Twisted Sector}

 Let us first explore   the large $N$ limit on the superconformal index for
$O(2N)$ $\mathcal{N}=4$ SYM with  $m$ hyper-multiplets in $2N$. The difference
between $O(2N)$ and $SO(2N)$ gauge group come from the `baryonic'
operators $K$ and $L$ of the form
\begin{align}
K &= \epsilon_{i_1 i_2  \ldots i_{2N}} A^{i_1}A^{i_2} \ldots
A^{i_{2N}}. \label{fund-baryon} \\
L &= \epsilon_{i_1 i_2 \ldots i_{2N}} \Phi^{i_1 i_2} \Phi^{i_3 i_4} \ldots \Phi^{ i_{2N-1} i_{2N}} \label{adj-baryon}
 \end{align}
$A^{i}$ denote the scalar in a
hyper-multiplet in $2N$,  $\Phi^{ij}$ in the adjoint representation.
These `baryonic' operators are invariant under $SO(2N)$ but
variant under $O(2N)$. Index contributions from these operators
start at $o(x^{N/2})$ which is negligible in the large N limit.
Thus, in the large N we don't need to distinguish $O(2N)$ gauge
group from $SO(2N)$.

%
%
%
%
%
%
 Superconformal index formula for the $SO(2N)$ SYM theory and its large N limit  are explicitly presented in the appendix \ref{largeN-bcd}. Large N  superconformal index from zero charge monopole, $I^{(0)}_{O(\infty):m}$,  can be expressed in terms of  plethystic expansion as
\begin{align}
I^{(0)}_{O(\infty):m} (x) 
&= \exp \big{[} \sum_{n=1}^\infty \frac{1}n I^{(0)}_{O(\infty):m:sp}(x^n)\big{]}, \nonumber
\end{align}
where the single particle index is given in eq.~\eqref{no-monopole-so2n} as
\begin{align}
I^{(0)}_{O(\infty):m:sp}(x)=&
\frac{x(3+2x^{1/2} + 2x - 2x^{5/2} -x^3)}{(1-x^2)^2} \nn \\
\quad & +(2m^2 +m) \frac{x}{(1-x)^2}+(2m^2 - m -1) \frac{ 2 x^{3/2}}{ (1-x)^2 (1+x)} \ .
\end{align}
On the other hand, 
the corresponding gravity bulk index is obtained in eq.~\eqref{I(0) on S7/<alpha,beta>}
\begin{align}
I^{(0)}_{(S^7/\mathbb{Z}_m)/\mathbb{Z}_2:sp} (x) =  \frac{x(3+2 x^{1/2} +2 x - 2x^{5/2}
-x^3)}{(1-x^2)^2},
\end{align}
after a careful analysis on the suitable ${\mathbb Z}_2$ modding of the gravity index on $AdS_4 \times (S^7/\mathbb{Z}_m)/{\mathbb Z}_2$.
As happened in $U(N)$ case,
twisted sector comes only from zero monopole charge sector.
 The twisted index is given by
\begin{align}
I_{O(\infty):m:twisted:sp} (x) & = I^{(0)}_{O(\infty):m:sp}(x) -
I^{(0)}_{(S^7/\mathbb{Z}_m)/\mathbb{Z}_2:sp} (x)  , \nonumber
\\
&=(2m^2 + m) A(x) + (2m^2 - m -1) B(x). \label{twisted of O(2N)}
\end{align}
Where $A(x), B(x)$ are given by
\begin{align}
A(x) = \frac{x}{(1-x)^2}, \quad B(x) = \frac{2x^{3/2}}{(1-x)^2
(1+x)}.
\end{align}
They are provided in \cite{Imamura:2009hc} by analyzing 7-d SYM  on $AdS_4\times S^3$.
The index for the 7d $U(1)$ SYM is
\begin{align}
I_{\textrm{7d  $U(1)$ SYM}} (x) = A(x) + B(x).
\end{align}
Note that $A(x)$ denotes the spectrum with the integer value while
$B(x)$ represents spectrum with half-integer value.\footnote{$A(x)$
is the index from states in the 7d SYM theory with $\beta=1$ and
$B(x)$ is the index from states with $\beta=-1$. $\beta$ is the
generator of $\mathbb{Z}_2$ in $AdS_4 \times S^3/\mathbb{Z}_2$,
which has the singular locus of the gravity background.} The above
expression of the twisted sector implies that we keep the adjoint of
the gauge group $Sp(2m)$ on D6 branes for the integer spectrum while
keeping antisymmetric representation for half-integer spectrum. Note
that the antisymmetric  representation of $Sp(2m)$ is made of a
irreducible representation of dim $m(2m-1)-1$   and one singlet. The
lowest integer spectrum represents the the gauge degrees of freedom
on $AdS_4$ so this must be the adjoint representation of dim
$m(2m+1)$ for $Sp(2m)$, which is the gauge group of D6s in the brane
setup. The factor $(2m^2 \pm m)$ in twisted sector index can also be
understood by explicitly constructing local operators  in
$\mathcal{N}=4$ $O(2N)$ SYM theory. Let $(A_I, B_I )|_{I=1,\ldots,
m} \in (2N, \bar{2N}) $ be $m$ fundamental hypermultiplets and
$(\Phi_1 , \Phi_2)$ be an adjoint hyper-multiplet multiplet in the
SYM. Since $2N = \bar{2N}$, combing $A_I$ and $B_I$ we define $2m$
fundamental chiral multiplets $F_I|_{I=1,\ldots, 2m}$ as ($I=1,
\ldots 2m$)
\begin{align}
F_I= A_I, \quad F_{2m+I} = B_I.
\end{align}
Then lower power in twisted sector index come from
\begin{align}
&F^{T}_{(I} F_{J)} \; : \; (2m^2 + m )x, \nonumber
\\
&F_{[I}^T \Phi_{1} F_{J]}, \; F_{[I}^T \Phi_{1} F_{J]} \; : \; 2(2m^2 - m )x^{3/2}.
\end{align}
Note that $F_I^T \Phi_{1,2} F_J = - F_J^T \Phi_{1,2} F_I$ due
to the property $\Phi_{1,2}^T = - \Phi_{1,2}$. F-term equation for
$\Phi$ (adjoint chiral multiplet in $\mathcal{N}=4$
vector-multiplet) will kill one combination of local operators of
the form $F_{[I}^T \Phi F_{J]}$ and give correct factor $(2m^2 -m
-1)$.

One can also compare the non-zero monopole charge part of field theory large N index, $I’_{O(2N):m}(x)$, and its corresponding SUGRA index,
$I’_{(S^7/\mathbb{Z}_m)/\mathbb{Z}_2}(x)$, and find exact match using the similar trick used in $U(2N)$ case.

Now let us turn our attention to the  $Sp(2N) \times SO(2m)$ case, the similar
analysis gives the twisted sector as, see eq.~\eqref{I(0) on S7/<alpha,beta>} and \eqref{no-monopole-sp2n}
\begin{align}
I_{Sp(\infty):m:twisted:sp} (x) &= I^{(0)}_{Sp(\infty):m:sp} (x) - I^{(0)}_{(S^7/\mathbb{Z}_m)/\mathbb{Z}_2:sp} \nn
\\
&= (2m^2 - m) A(x) + (2m^2 + m -1) B(x) \; .
\end{align}
Again this has the simple interpretation in terms of the gauge group $SO(2m)$ for D6 branes. The integer spectrum has
the adjoint representation of dim $ m(2m-1)$ for $SO(2m)$,
while the half integer spectrum has the traceless symmetric representation of $SO(2m)$. 
 One can also find operators in the $\mathcal{N}=4$ $Sp(2N)$ SYM
corresponding to twisted sector index.
\begin{align}
&F_{[I}^T \mathbb{J} F_{J]}\; : \; (2m^2 -m) x, \nonumber
\\
&F_{(I}^T \mathbb{J}\Phi_{1} F_{J)},\; F_{(I}^T \mathbb{J}\Phi_{2} F_{J)} \; : \; 2 (2m^2 +m) x^{3/2}.
\end{align}
$\mathbb{J}$ denote  the skew-symmetric  form of  $Sp$-group.
 Note that $F_I^T \mathbb{J}F_J = - F_J^T
\mathbb{J}F_I$ due to the property $\mathbb{J}^T = - \mathbb{J}$ and
$F^T_I \mathbb{J}\Phi_{1,2} F_J = F^T_J \mathbb{J}\Phi_{1,2} F_I$
due to the property $\mathbb{J}\Phi_{1,2} +\Phi_{1,2}^T \mathbb{J}
=0$. All these are consistent with the orientifold projection $SU(2m)
\rightarrow Sp(2m)/SO(2m)$.

Finally for the $SO(2N+1) \times Sp(2m)$
case, the twisted sector contribution is given by eq. (\ref{twisted
of O(2N)}) as it should be since twisted sectors are coming from
D6 strings, which has the same $Sp(2m)$ projection for both
$O(2N)$ and $SO(2N+1)$ theories.

The index computation strongly suggests that in the large N limit,
the IR superconformal theories of ${\cal N}=8$ $O(2N), Sp(2N),
SO(2N+1)$ super Yang-Mills theory are dual to the gravity theory on
$AdS_4 \times S^7/{\mathbb Z}_2$. As the ABJ(M) models of the
$U(N)_2\times U(N)_{-2}$ and $U(N)_2 \times U(N+1)_{-2}$ groups
 also have the same gravitational dual. In order to tell which theory is mapped to
which one, we need the information on finite N. Such information is
given by ${\mathbb Z}_2$ orbifolded theory.  In the Type IIB
Hanany-Witten setup with D3/NS5/ (1,2) 5brane , introducing $m$ NS
5branes gives rise to ${\mathbb C}^2/{\mathbb Z}_m$ orbifold
singularities. If we consider 2 NS5 branes, this also leads to
$U(N)_0$ factor in addition to $Z_2$ orbifold. We already saw that
$O(2N)$ SYM with one fundamental hyper is mapped to $U(N)_2\times
U(N)_{-2} \times U(N)_0$ SCSM while $USp(2N)$ SYM is mapped to
$U(N)_2\times U(N+1)_{-2} \times U(N)_0$ SCSM. Since we can take the
large N limit for these theories, this tells us that for ${\mathbb
Z}_2$ invariant sectors we have to identify $O(2N)$ theory with
$U(N)_2\times U(N)_{-2}$ and $USp(2N)$ theory with $U(N+1)_2\times
U(N)_{-2}$. Hence this identification is natural for the whole
theory. Thus our claim is that ${\cal N}=8$ $ O(2N)$ SYM flows to
$U(N)_2\times U(N)_{-2}$ ABJM and ${\cal N}=8$ $ SO(2N+1), Sp(2N)$
SYM flows to $U(N+1)_2 \times U(N)_{-2}$ ABJ theory. In the case of
$Sp(2N)$ we are assuming $\vartheta=0$ in (\ref{basicaction}).

Now consider the Higgsing  pattern of ABJ(M) theory. Starting from
$U(N)_{2}\times U(N)_{-2}$ ABJM this can be Higgesed down to
$U(1)_2\times U(1)_{-2}\times U(1)^{N-1}$ where the last factor
represents the product of the $U(1)$ Yang-Mills theory. This can be
achieved by giving the scalar vev of ABJM theory
\begin{equation}
Z_4=\bar{Z}^4=\left( \begin{array}{ccc}
                0 & 0 & 0\cdots \\
               0 & R_1 & 0 \cdots \\
               0 & 0 & R_2 \cdots    \end{array}  \right).
\end{equation}
We already saw that $U(1)_2\times U(1)_{-2}$ ABJM theory is the IR
limit of $O(2)$ or $Sp(2)$ with $\vartheta=\pi$. And the Higgsing
pattern is $O(2N) \rightarrow O(2)\times U(1)^{N-1}$ and
$Sp(2N)\rightarrow Sp(2) \times U(1)^{N-1}$. We conclude that the IR
limit of ${\cal N}=8 \, O(2N), Sp(2N)$ with $\vartheta=\pi$ is given
by $U(N)_2\times U(N)_{-2}$ ABJM theory. For $U(N)_{2} \times
U(N+1)_{-2}$ ABJ theory, it is Higgsed to $U(1)_2\times U(2)_{-2}
\times U(1)^{N-1}$ where the last factor represents the product of
the $U(1)$ Yang-Mills theory. Since the IR limit of $SO(3)$ and
$Sp(2)$ with $\vartheta=0$ is given by $U(1)_{2} \times U(2)_{-2}$
ABJ theory, the IR limit of $SO(2N+1)$ and $Sp(2N)$ with $\theta=0$
should be given by $U(N)_{2} \times U(N+1)_{-2}$ ABJ theory.

\section{Mirror symmetry and ${\cal N}=4$ Supersymmetric Chern-Simons theory}

\subsection{Basic setup and the index computation}

In three-dimensions, there are important classes of ${\cal N}=4$
superconformal field theories which are extensively discussed in
the context of mirror symmetry. These can be described in
Hanany-Witten setup as a collection of D3/NS5/D5 branes. The
mirror symmetry is realized as  the S-dual transformation of
$SL(2,Z)$ of Type IIB theory, which interchanges NS5 and
D5\footnote{For later purpose, we had better fix the world colume
directions of D3/NS5/D5 branes. D3 spans (0123), D5 spans (012456)
and NS5 spans (012789)}. In the field theory side, this interchanges
Coulomb and Higgs branch. At the origin of the moduli space where
the Coulomb and the Higgs meet, we have 3d superconformal field
theory. The natural question is if these superconformal field theories
admit Chern-Simons type description \cite{HLLLP, ImamuraN=4}. The
answer is positive and the subsequent computation gives an
impressive confirmation.

\begin{figure}[htbp]
   \centering
   \includegraphics[scale=0.6]{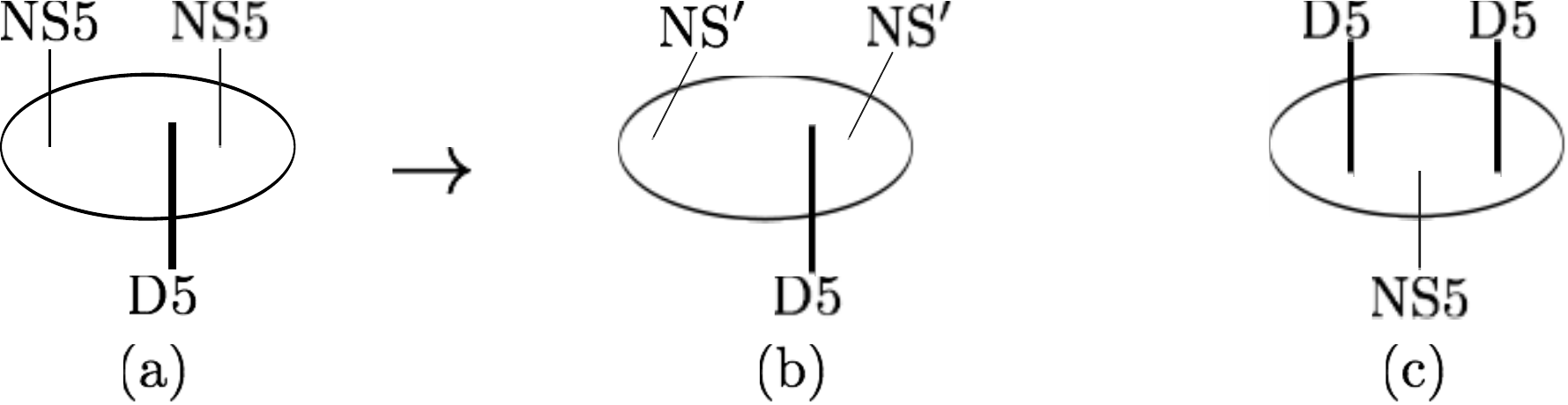} 
 \caption{ (a) A configuration of $N$ D3s, 2 NS5s, and one D5. (b) T-dual transformation of (a). (c) S-dual transformation of (a). }
    \label{fig1}
\end{figure}

 Let us start with the simplest example, $N$
D3 branes with 2 NS5s and one D5 as shown in Fig~\ref{fig1} (a).
This has ${\cal N}=4$ supersymmetry in 3-d and we have $U(N) \times
U(N)$ YM with hypermultiplets transforming $(N, \bar{N}),
(\bar{N},N), (1,N)$. If we take the T-dual transformation
$\tau\rightarrow \tau+1$ of $SL(2,Z)$ in Type IIB setting, D5 branes
are invariant while NS5 branes are turned into (NS5, D5)=(1,1)
brane, which we will call NS' brane subsequently. These are the
configurations considered by Imamura and Kimura \cite{Imamura:2008nn}. We have ${\cal N}=4$ SCSM
 with the gauge group $U(N)_1\times U(N)_0\times
U(N)_{-1}$ with bi-fundamental hypers in
\begin{equation}
(N,\bar{N},1),(N,\bar{N},1),(\bar{N},1,N).
\end{equation}
 The
subscript in the gauge group denotes Chern-Simons level. This is
shown in Fig~\ref{fig1} (b). This suggests that the ${\cal N}=4$
SYM of Fig~\ref{fig1} (a) flows to the
SCSM of Fig~\ref{fig1} (b). Alternatively, one can
take S-dual transformation from Fig~\ref{fig1} (a) to obtain
Fig~\ref{fig1} (c), exchanging NS5 and D5 branes. The resultant
theory is $U(N)$ YM  with two fundamental and one adjoint
hyper-multiplets. The index computation confirms this.\footnote{For  ${\cal N}=4$ abelian theories  related
by mirror symmetry, one has analytic proof for the equality of the index at \cite{Spiridonov}. It would be interesting
to find the similar proof for the cases handleded in this paper.} For example,
one obtains
  \begin{align}
&   I^{ U(1) \times U(1) }_{YM \oplus 1  fund.hyper}  \  =      I^{U(1) }_{YM  \oplus 2 fund. hypers }  \
 =   I^{ U(1)_1 \times U(1)_0 \times U(1)_{-1}  }_{ SCS} \nn \\
&   =1 + 2 x^{1/2} + 9 x + 14 x^{3/2} + 2 2 x^2 + 20 x^{5/2} +
25 x^3 + 34 x^{7/2} + 62 x^4 + 74 x^{9/2} + O(x^5)  . \nn
 \end{align}

 \begin{figure}[htbp]
    \centering
    \includegraphics[scale=0.6]{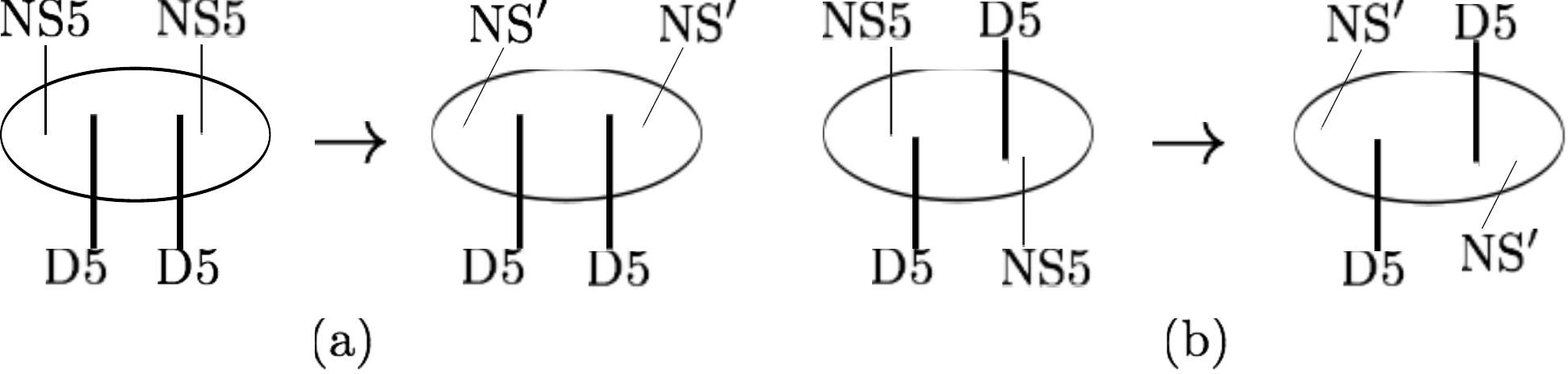} 
    \caption{Two possible configurations of 2 NS5s and 2 D5s }
    \label{fig2}
 \end{figure}

Now consider another example where now we have 
2 NS5s and 2 D5s. Note that these are self-mirror configurations. As
shown in Fig ~\ref{fig2} (a) (b) there are two possible
configurations of NS/D5. One is NS-NS-D5-D5 as appearing in Fig
~\ref{fig2} (a) and the other is NS-D5-NS-D5 as appearing in Fig
~\ref{fig2} (b).  In (a), the first one gives $U(N) \times U(N)$ Yang-Mills
with hyper-multiplets in $(N,\bar{N}), (\bar{N}, N), 2(1,N)$. Under
the T-dual transformation it turns into Chern-Simons quiver theory
$U(N)_1\times U(N)_0 \times U(N)_{-1} \times U(N)_{0}$ with
bi-fundamental hyper-multiplets in
\begin{equation}
(N,\bar{N},1,1), (1,N, \bar{N},1),(1,1,N, \bar{N})( \bar{N},1,1, N). \label{bi-fund-rep}
\end{equation} The index computation gives the same result. As an
example we exhibit the index with $N=1$
 \begin{align}
&  I^{U(1)_1 \times U(1)_0 \times U(1)_{-1} \times U(1)_0 }_{CS}=
I^{U(1) \times U(1)}_{  YM } = 1+ 12 x + 42 x^2 + 48x^3 + 115 x^4 +
188 x^5 + O(x^6).   \label{index-1}  \end{align}

In Fig ~\ref{fig2} (b), the first one gives $U(N)\times U(N)$ Yang-Mills theory with hyper-multiplets in $(N,\bar{N}), (\bar{N},N),
 (N,1), (1,N)$. Under the T-dual transformation, it is mapped to
 $U(N)_1\times U(N)_{-1}\times U(N)_1\times U(N)_{-1}$ quiver
Chern-Simons theory with bi-fundamental hyper-multiplets in \eqref{bi-fund-rep}. Again for $N=1$, the index computation gives
\begin{align} & I^{U(1)_1 \times U(1)_{-1} \times U(1)_1 \times U(1)_{-1}}_{CS}  = I^{U(1) \times U(1)}_{YM^{ \prime}}  \nn \\
& \quad
= 1+ 8x + 8x^{3/2} + 18 x^2 + 16 x^{5/2}+28 x^3 + 63 x^4 + 80 x^{9/2} + 56 x^5 + O (x^{11/2})
\label{index-2}
\end{align}
where now the hyper-multiplets in the YM$^{\prime}$ have the charges $(1,-1) \oplus (-1,1) \oplus (1,0) \oplus (0,1)$ under the gauge group, in contrast to that the hyper-multiplets in YM in \eqref{index-1} have charges $ (1, -1) \oplus (-1,1) \oplus 2 ( 1, 0)$.

\begin{figure}[htb]
   \centering
   \includegraphics[scale=0.6]{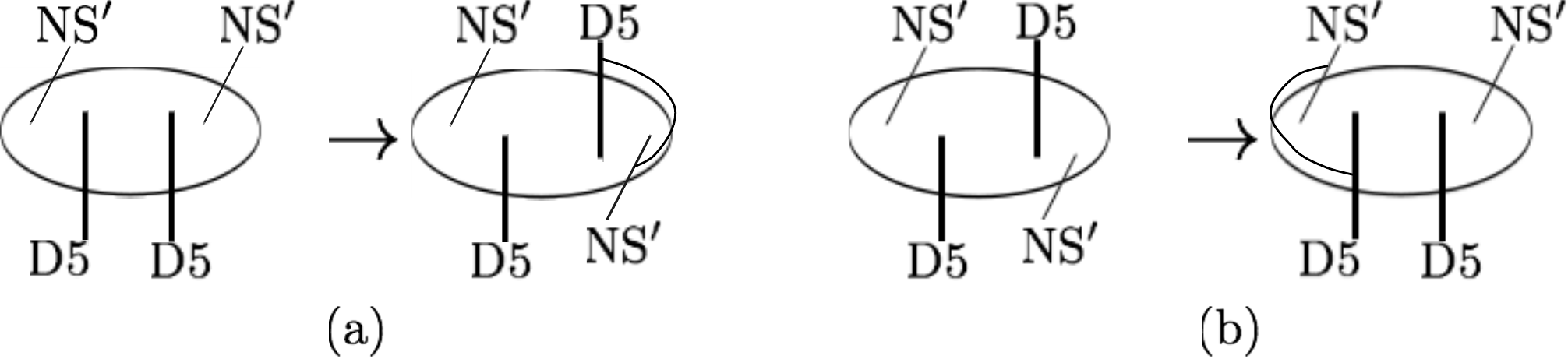} 
   \caption{D3-brane creation due to Hanany-Witten effect }
   \label{fig3}
\end{figure}

The interesting feature is that they have the same moduli space yet
different index. One can easily see why this is so. When one tries to exchange NS and D5 (or NS' and D5)
to obtain one brane configuration from the other, we create D3 brane
between NS' and D5 \cite{Hanany:1996ie}. This suggests an
equivalence among ${\cal N}=4$ Chern-Simons quiver theories
\begin{equation}
U(N)_1\times U(N)_0\times U(N)_{-1}\times U(N)_0 \sim U(N+1)_1\times
U(N)_{-1}\times U(N)_1\times U(N)_{-1}
\end{equation}
 as appearing in Fig~\ref{fig3}.
Indeed the index computation confirms this, for example $N=1$ case
gives,
\begin{align}
I^{U(1)_1 \times U(1)_{-1} \times U(1)_{1} \times U(1)_{-1} }_{CS} (x) &=I^{U(2)_1 \times U(1)_0 \times U(1)_{-1}
\times U(1)_0 }_{CS} (x) \nn \\
I^{U(1)_1 \times U(1)_{0} \times U(1)_{-1} \times U(1)_0}_{CS} (x) & =
I^{U(2)_{-1} \times U(1)_1 \times U(1)_{-1} \times U(1)_1 }_{CS}  (x) \nn
\end{align}
where the explicit forms are given in eq.~\eqref{index-1} and~\eqref{index-2}.

For Chern-Simons theories with higher level, similar brane creation effect can
occur. Taking account of this, one can see the equivalence of
various theories, which can be regarded as the ${ \cal N}=4$ generalization of the ${ \cal N}=6$ dualities in \cite{Aharony:2008gk}. For example, the following two ${\cal N}=4$ SCSMs are expected to be equivalent,
 \begin{align}& U(N)_{k}  \times U(N)_{-k} \times U(N)_{k} \times U(N)_{-k}
 \sim  U(N+|k|)_k \times U(N)_0 \times U(N)_{-k} \times U(N)_0 \ .   \nn \end{align}
An explicit example for $N=1, k=2$ is
\begin{align}
I^{U(1)_2 \times U(1)_{-2} \times U(1)_2 \times U(1)_{-2} }_{CS} =
I^{U(3)_2 \times U(1)_0 \times U(1)_{-2} \times U(1)_0 }_{CS} & = 1
+ 4x +18 x^2 + 16 x^3 + 35 x^4 + O (x^5). \nn \end{align}

Now the pattern for general configurations is obvious. In the
appendix we look for the partition function of these theories and
found agreements. We also turn on FI and mass parameters and show
how these are mapped under the dualities between SYM
and SCSM and provide subsequent
interpretation in the next subsection.

\subsection{FI and mass parameters}

 In this subsection,  we generalize the result by adding mass and FI parameters.
 The following discussion includes the ABJM and ${ \cal N}=8$ SYM considered in \cite{Kapustin:2010xq}
 as a special case $m=n=1$, where $m$ ($n$) is the number of D5's (NS5 or NS$^{\prime}$'s).
Since the FI and mass parameter do not carry the color index, we will
consider the map of parameters of abelian theories.

\begin{figure}[htbp]
   \centering
   \includegraphics[scale=0.6]{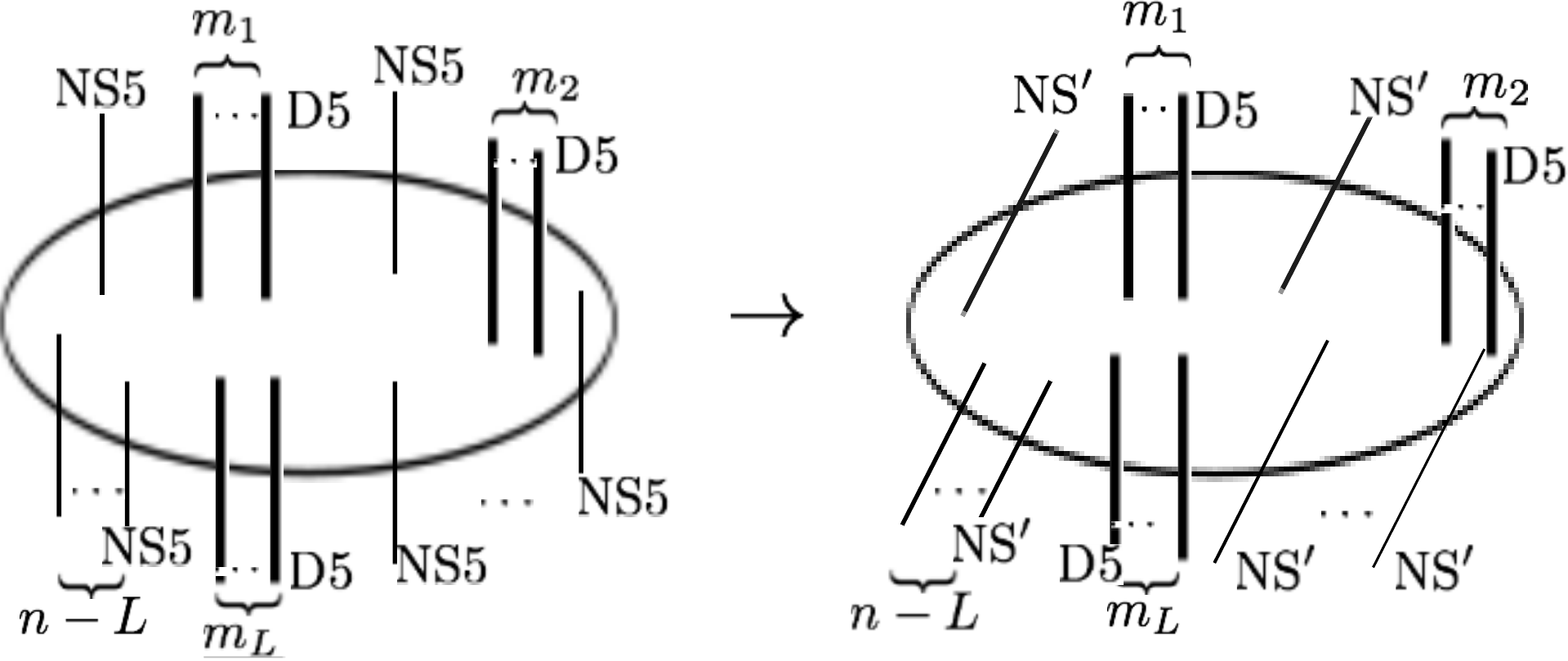} 
   \caption{(a) A generic Hanany-Witten set up. (b) T-dual transformation from (a)} \label{fig4}
\end{figure}

As the first example, we consider the case that one NS5 and one D5 are
alternating, i.e. $m=n=L, \ m_1 = \ldots = m_n = 1,$ in the
Fig.~\ref{fig4}. The parameters of interest are
\begin{align}
\begin{array}{cclr}
\mbox{YM} & U(1)^n & \mbox{bi-fundamental masses } \omega_I & \qquad  ( I = 1 , \ldots, n)  \\
& & \mbox{fundamental masses } \mu_I  &  \\
& & \mbox{FI parameters } \eta_I &   \\
\mbox{CS} & \quad U(1)^{2n} \quad  & \mbox{bi-fundamental masses } \xi_I  & \qquad (I = 1, \ldots, 2n)   \\
& & \mbox{FI parameters } \zeta_I &   \\
\end{array}
\label{parameters}
\end{align}
The Chern-Simons level is alternating $1$ and $-1$, $U(1)_1 \times U(1)_{-1} \times \ldots \times U(1)_{-1}$.

Some parameters in \eqref{parameters} are not independent  \cite{Kapustin:2010xq}. One can see it from the partition function. If we turn on all parameters, the partition
functions are changed to
\begin{align}
Z_{YM} ( \omega, \mu, \eta ) & = \int ( d \sigma)^n \frac{e^{ 2 \pi
i \sum_{I=1}^n \eta^I \sigma^I }}{ \prod_{I=1}^n  \cosh ( \pi (
\sigma^I - \sigma^{I+1} + \omega_I)) \prod_{I=1}^n \cosh ( \pi (
\sigma^I + \mu^I ) )},
\nn  \\
Z_{CS} ( \xi, \zeta ) & = \int ( d \sigma)^{2n} \frac{ e^{ \pi i ( (
\sigma^1)^2 - ( \sigma^2)^2 + \cdots + ( \sigma^{2n-1})^2 - (
\sigma^{2n})^2 )+ 2 \pi i \sum_{I=1}^{2n} \zeta^I \sigma^I }}{
\prod_{I=1}^{2n} \cosh ( \pi ( \sigma^I - \sigma^{I+1} + \xi_I ) )
} . \nn
 \end{align}
However, by the constant shift of integral variables, one can absorb one of the
fundamental masses of YM.  In the same way, one can set the bi-fundamental masses of YM and CS to
be same respectively,
\begin{align}
& \omega_1 = \omega_2 = \cdots = \omega_n : = \omega,
\qquad \xi_1 = \xi_2 = \ldots = \xi_{2n} := \xi . \label{bi-fund-mass} \end{align}
The FI parameters of CS can be set to satisfy
\begin{align}
\sum_{I=1}^n \zeta^{2I-1} =  \sum_{I=1}^{n} \zeta^{2I}   : = n \zeta
\label{zeta} \end{align} for some fixed $ \zeta$, which becomes obvious in \eqref{zeta-cond}
after changes of variables . Thus the number of independent
parameters are  $2n $; $1$ bi-fundamental mass, $n-1$ fundamental masses, and $n$ FI
parameters in YM; $1$ bi-fundamental mass and $2n-1$ FI
parameters in CS.

Imposing \eqref{bi-fund-mass}, the partition functions become
\begin{align}
Z_{YM} ( \omega, \eta) & = \int ( d \sigma)^n \frac{e^{ 2 \pi i
\sum_{I=1}^n \eta^I \sigma^I }}{ \prod_{I=1}^n  \cosh ( \pi (
\sigma^I - \sigma^{I+1} + \omega)) \prod_{I=1}^n \cosh ( \pi (
\sigma^I + \mu^I ) )}, \label{ym-final}
\\
Z_{CS} ( \xi, \zeta ) & = \int ( d \sigma)^{2n} \frac{ e^{ \pi i ( (
\sigma^1)^2 - ( \sigma^2)^2 + \cdots + ( \sigma^{2n-1})^2 - (
\sigma^{2n})^2 )+ 2 \pi i \sum_{I=1}^{2n} \zeta^I \sigma^I }}{
\prod_{I=1}^{2n} \cosh ( \pi ( \sigma^I - \sigma^{I+1} + \xi ) )  }. \label{cs-temp}
\end{align}
One can use formulas in section \ref{useful} to rewrite $Z_{CS} ( \xi, \zeta)$ as
\begin{align}
Z_{CS}( \xi, \zeta)
& = \int ( d \sigma)^{2n} ( d \tau)^{2n}  \frac{ e^{ \pi i ( ( \sigma^1)^2 - ( \sigma^2)^2 + \cdots + ( \sigma^{2n-1})^2 - ( \sigma^{2n})^2 )+ 2 \pi i \sum_{I=1}^{2n} \left(   \tau^I \xi  +  \sigma^I ( \tau^I - \tau^{I-1}  + \zeta^I) \right) }}{ \prod_{I=1}^{2n} \cosh ( \pi \tau^I)  } \nn \\
& = \int ( d \tau)^{2n} \frac{ e^{ 2 \pi i ( \sum_{a = 1}^n ( \tau^{2a +1}- \tau^{2a -1} )  \tau^{2a} + 2 \pi i \sum_{a=1}^n  \left( \tau^{2a} ( \zeta^{2a} + \zeta^{2a+1} + \xi ) - \tau^{2a -1} ( \zeta^{2a-1} + \zeta^{2a} - \xi ) \right) } }{ \prod_{a=1}^{n} \cosh ( \pi \tau^{2a}) \prod_{a=1}^{n} \cosh ( \pi \tau^{2a -1} )} \label{zeta-cond} \\
&= \int ( d \kappa)^n \frac{ e^{  2 \pi i \sum_{I=1}^n \kappa^I (
\zeta^{2I-1} + \zeta^{2I } - \xi) }}{ \prod_{I=1}^n \cosh ( \pi
\kappa^I ) \cosh ( \pi ( \kappa^{I} - \kappa^{I+1} + \zeta^{2I} +
\zeta^{2I+1} + \xi ) )} \nn
\\
&=\int ( d \tau)^n  \frac{ e^{  2 \pi i \sum_{I=1}^n \tau^I (
\zeta^{2I-1} + \zeta^{2I } - \xi) }}{ \prod_{I=1}^n  \cosh ( \pi ( \tau^{I} - \tau^{I+1} + \mu^I - \mu^{I+1} + \zeta^{2I} +
\zeta^{2I+1} + \xi ) ) \cosh ( \pi
(\tau^I + \mu^I) ) } .
\label{cs-final}
\end{align}
In the last two lines, we define $ \kappa^I = - \tau^{2I-1}$ then recycle $\tau$ for $ \tau^I := \kappa^I + \mu^I$. The equalities hold up to an overall phase. Comparison of \eqref{ym-final} and \eqref{cs-final} gives the following map between mass and FI parameters \begin{align}
\omega & = ( 2 \zeta + \xi) \nn \\
\mu^I - \mu^{I+1} & = 2 \zeta - ( \zeta^{2I} + \zeta^{2I+1}),  \qquad (I = 1, \ldots, n)
\nn \\
\eta^I & =  ( \zeta^{2I-1} + \zeta^{2I} - \xi) , \qquad (I = 1,
\ldots, n) .  \end{align} For $(n,m)=(1,1)$, it reduces to ${ \cal N}=8$ SYM/
ABJM result in \cite{Kapustin:2010xq},
\begin{align}
\omega = 2 \zeta + \xi, \qquad \eta = 2 \zeta - \xi.
\label{abjm-sym}
\end{align}

The peculiar feature is that in the mapping from YM type to SCSM
theory, FI term and  mass term are mixed up with each other. In the
Yang-Mills theory, there's $SO(4)=SU(2)_L\times SU(2)_R$ R symmetry
denoted by $R_{YM}$. Under this, FI term transforms as (1,3) and
mass term transforms as (3,1). In the Hanany-Witten setup,
fundamental mass term is given by the transverse location $x_4, x_5,
x_6$ of D5 branes while FI term is the transverse location $x_7,
x_8, x_9$ of NS5 branes. Now in the brane setup it's obvious that
among $SU(2)_L\times SU(2)_R$ $ R_{YM}$ symmetry only its diagonal
combination survives since the NS' brane is rotated in the 456-789
planes with respect to NS brane. Hence it's natural that FI term and
mass term in the YM theory is mixed up in the SCSM setting.  The
$SO(4)_R'=SU(2)_L'\times SU(2)_R'$ symmetry in the SCSM theory of
interest arise where the R-symmetry transformation on the
hypermultiplets and the twisted hypermultiplets are different, which
cannot be seen in SYM theory. Note that we have $2n-1$ FI parameters
which should transform as $(3,1)$ or $(1,3)$ under
$SO(4)_{R'}=SU(2)_{L'}\times SU(2)_{R'}$. The distinction is not
important since the interchange of $SU(2)_{L'}$ and $SU(2)_{R'}$
interchanges hypermultiplets and twisted hypermultiplets, which
leads to the same theory.

As the second example,  we consider the case that all D5
branes are between the first and the second NS5-branes. In
Fig.~\ref{fig4}, it corresponds to $L=1, m_1=m $. The parameters of interest are
\begin{align}
\begin{array}{cclr}
\mbox{YM} & U(1)^n & \mbox{bi-fundamental mass } \omega_I & \qquad  ( I = 1 , \ldots, n)  \\
& & \mbox{fundamental mass } \mu_a  &  (a = 1, \cdots, m) \\
& & \mbox{FI parameters } \eta_I & ( I = 1, \ldots, n)    \\
\mbox{CS} & \quad U(1)^{n+m} \quad  & \mbox{bi-fundamental mass } \xi_I  & ( I = 1, 2, \ldots, n+m)     \\
& & \mbox{FI parameters } \zeta_I &  \\
\end{array}
\nn
\end{align}
The Chern-Simons level is $\vec{k}=(1,0, \ldots, 0, -1, 0, \ldots, 0)$, nonvanishing for the first and the $(m+1)$th gauge group. Again, not the all
parameters are independent. The constant shift of integral
variables in the partition function can impose
\begin{align}
& \omega_1 = \omega_2 = \cdots = \omega_n : = \omega, \qquad  \mu_1 = 0 ,  \nn \\
& \xi_1 = \xi_{m+1} : = \xi, \qquad \xi_I = 0 \quad
\mbox{for }I \neq 1, m+1, \qquad \zeta_1 = \zeta_{m+1} : =\zeta \nn
\end{align}
Thus the number of independent parameters is $(n+m)$; $1$ bi-fundamental mass, $m-1$ fundamental mass, and $n$ FI
parameters in YM;  $1$ bi-fundamental mass and $n+m-1$ FI
parameters in CS.

  We now repeat the derivation. The Chern-Simons partition function can be written as an integral over $(n+m)$ variables,
 \begin{align}
Z_{CS} ( \xi, \zeta) = \int ( d \sigma)^{n+m} \frac{ e^{ \pi i (( \sigma^1)^2 - ( \sigma^{m+1})^2 )+ 2 \pi i \sum_{I=1}^{n+m} \zeta^I \sigma^I  } }{ \prod_{I=1}^{n+m} \cosh ( \pi ( \sigma^I - \sigma^{I+1} + \xi ))}.
  \nn \end{align}
   We use \eqref{fourier-2} to introduce $(n+m)$  $ \tau$-variables to Chern-Simons partition function, then integrate out $ \sigma$ and $ \tau$ variables in turn except $ \sigma^1, \sigma^{m+1}, \tau^1, \tau^{m+1}$. The Gaussian integration of $ \sigma^1$ and $ \sigma^{m+1}$ results in
\begin{align}
Z_{CS} ( \xi, \zeta_I) & =  \int  ( d \tau^1 d \tau^{m+1} )  \frac{ 1 }{   \cosh
( \pi \tau^1 ) \cosh ( \pi \tau^{m+1} )
   }
\nn \\
& \qquad \times \frac{e^{ 2 \pi i  (   (- \xi+ ( \zeta_1 +
\zeta_{m+1}) ) \tau^1+  ( \xi+( \zeta_1+\zeta_{m+1} ) ) \tau^{m+1}
)}}{  (   \prod_{I=2}^{m}  \cosh  \pi ( \tau^1 + \sum_{
\alpha=2}^{I}  \zeta_{ \alpha}  )  ( \prod_{a=2}^{n}   \cosh \pi (
\tau^{m+1} - \sum_{\alpha=m+2}^{m+a} \zeta_{ \alpha} ) ) }.
\label{cs-fi}
\end{align}
On the other hand, the YM partition function is given by
\begin{align}
 Z_{YM} ( \vec{\eta}, \omega, \vec{ \mu})  & = \int ( d \sigma)^n  \frac{e^{ 2 \pi i  \sum_{I=1}^n \eta_I \sigma^I }  }{  \prod_I \cosh ( \pi (  \sigma^I - \sigma^{I+1} +\omega ) ) ) \cosh ( \pi \sigma^1 ) \prod_{ a=2}^{m} \cosh ( \pi (\sigma^1 + \mu_a ) ) }. \nn \end{align}
 We use \eqref{fourier-2} to introduce $n$ $ \tau$-variables, then integrate out $ \sigma$ and $ \tau$ variables in turn except $ \sigma^1, \tau^1$. It results in
 \begin{align}
Z_{YM} (\vec{ \eta}, \omega, \vec{ \mu})
& = \int d \sigma d \tau \frac{e^{ 2 \pi i ( \sum_{ I=1}^{n} \eta_I
) \sigma  + 2 \pi i  (n \omega ) \tau  } }{ \prod_{I=2}^{n} \cosh
( \pi ( \tau - \sum_{\alpha=2}^{I} \eta^{\alpha} )) \cosh ( \pi \tau
)   \prod_{a=1}^{m} \cosh ( \pi (\sigma+ \mu_a) ) }.  \label{ym-fi}
\end{align}
 Then comparison of \eqref{cs-fi} and \eqref{ym-fi} gives the following map
\begin{align}
&  n \omega = 2 \zeta +\xi   , \qquad \sum_{I=1}^n \eta_I = 2 \zeta -\xi , \nn \\
&  \mu_a  =  \zeta_a ,  \quad ( a = 2, 3, \cdots, m) ,
\qquad  \eta_I = \zeta_{m+I}, \quad (I = 2, 3, \cdots, n)
 \end{align}
Again, for $(n,m)=(1,1)$, it reduces to the result of ${ \cal N}=8$
SYM/ ABJM in \eqref{abjm-sym}. Again we can observe the mixing
between $FI$ and mass terms.

\section{Concluding Remarks}

We found the equivalence  between the low energy superconformal
field theory of $\CN=8$ super Yang-Mills theories of classical
groups ABCD and the   superconformal ABJ(M) models of the
Chern-Simons level $k=1,2$. The supportive evidence is found from
the match of superconformal indices.

We also partially find the match in the partition function as shown in Appendix C.
However, we have not exhausted the calculation of the indices for the $\CN=8$ superconformal field theories. Besides
the IR limit of the SYM theories of $SU(N)$, $SO(2N)$ and exceptional gauge groups, there are also the
 Chern-Simons matter  model of $SU(N)_k\times SU(N)_{-k}$ with $k=1,2$ and
BLG model  of arbitrary $k$~\cite{BL,Gustavsson}, which has been studied extensively.
There may be more $\CN=8$ superconformal field theories besides what has been discussed.
It remains to be seen.

\section*{Acknowledgement}

We would like to thank Seok Kim for useful discussions.
  KL and JP are  supported in part by NRF-2005-0049409 through Sogang Univ. SRC CQUeST.
  KL  is supported in part by NRF  Korea Grants NRF-2009-0084601 and NRF-2006-0093850.
  JP is supported by the NRF Korea Grants R01-2008-000-20370-0 and No. 2009-0085995.
  DG is supported in part by Basic Science Research Program through the National Research Foundation of Korea(NRF) funded
        by the Ministry of Education, Science and
        Technology(2010-0007512).
  JP appreciates APCTP for its stimulating environment for
  research and acknowledges Simons summer workshop on mathematics and
  physics 2011 for hospitality while the paper is finalized.

\newpage

\appendix
\section{Index over gravitons in $AdS_4 \times S^7/\Gamma$} \label{appA}
In this section, we summarize relevant results on graviton index on $AdS_4 \times S^7/\Gamma$ with
discrete quotient $\Gamma$.
The index of single graviton in $AdS_4 \times S^7$ is given by \cite{Bhattacharya:2008zy}
\begin{align}
I_{S^7:sp} (x,y_1 , y_2 , y_3 ) =\Tr_{\textrm{(single graviton)}} (-1)^F x^{\epsilon_0 + j_3} y_1^{h_1} y_2^{h_2} y_3^{h_3}
= \frac{(\textrm{numerator})}{(\textrm{denominator})},
\end{align}
where
\begin{align}
(\textrm{numerator}) = &\sqrt{y_1 y_2  y_3} (1+ y_1 y_2 + y_2 y_3 + y_3 y_1 )x^{\frac{1}2}
- \sqrt{y_1 y_2 y_3} (y_1 + y_2 + y_3 + y_1 y_2 y_3)x^{\frac{7}2} \nonumber
\\
& + (y_1 y_2 + y_2 y_3 +y_3 y_1 + y_1  y_2  y_3 (y_1 + y_2 + y_3))(x^3 - x), \nonumber
\\
(\textrm{denominator}) = &(1-x^2) (\sqrt{y_3} - \sqrt{x y_1 y_2})(\sqrt{y_1} - \sqrt{x y_2  y_3})
(\sqrt{y_2} - \sqrt{x y_3  y_1})(\sqrt{y_1  y_2  y_3} - \sqrt{x}).
\end{align}
$\{h_i\}_{i=1,\ldots, 4}$ denote four Cartans of  $SO(8)$ isometry in $S^7$. $\epsilon_0$ and
$j_3$ are two Cartans of $SO(2,3)$ isometry in $AdS_4$ which are called energy and spin respectively.
The $SO(8)$ generators act on $\mathbb{C}^4 = \textrm{Cone}(S^7)$ as follows
\begin{align}
&h_1 = \textrm{diag}(\frac{1}2 , -\frac{1}2, -\frac{1}2 ,  \frac{1}2), \; h_2
= \textrm{diag}(\frac{1}2 , -\frac{1}2, \frac{1}2 , - \frac{1}2), \nonumber
\\
&h_3 = \textrm{diag}(-\frac{1}2 , -\frac{1}2,  \frac{1}2 ,  \frac{1}2), \; h_4
= \textrm{diag}(\frac{1}2 , \frac{1}2,  \frac{1}2 ,  \frac{1}2).
\end{align}
Only gravitons satisfying following BPS bound  contribute to the index.
\begin{align}
\epsilon_0 = j_3  +h_4.
\end{align}
The graviton index on $AdS_4 \times S^7/\Gamma$ can be obtained by keeping only the contribution form
$\Gamma$-invariant gravitons.

Consider the case when  $\Gamma = \mathbb{Z}_m$, whose generator
$\exp(\frac{2\pi i }{m}h)$ with $h = h_2 - h_1 $ act on the
$\mathbb{C}^4$ as
 \begin{align}
 h  = \textrm{diag} (0 , 0 , 1, -1).
 \end{align}
To keep contribution from $\mathbb{Z}_m$-invariant gravitons, it's convenient to introduce to chemical potential
$y$ for the charge $h$.
\begin{align}
I_{S^7:sp} (x, y ) &= I_{S^7:sp} (x, y_1 = 1/y, y_2 = y , y_3 =1), \nonumber
\\
& = \sum_{n \in \mathbb{Z}} I^{(n)}_{S^7;sp} (x) y^n,
\end{align}
where
\begin{align}
I^{(n)}_{S^7;sp} (x) =
\left \{
\begin{array}{ll}
\frac{2 (x^{1/2} + x - x^{5/2})}{1-x^{1/2}-x^2 + x^{5/2}}, & \hbox{$n=0$} \\
x^{|n|/2} \frac{(1+x^{1/2}+x)^2}{1-x^2}, & \hbox{ $n \neq 0$}. \label{I^{(n)}_{S^7;sp}}
\end{array}
\right.
\end{align}
Then, the graviton index in $AdS_4 \times S^7/\mathbb{Z}_m$ can be written as
\begin{align}
I_{S^7/\mathbb{Z}_m ; sp} (x,y)= I^{(0)}_{S^7 ;sp} + \sum_{n\neq 0 } y^{mn} I^{(mn)}_{S^7;sp} (x).
\end{align}

Consider the case when $\Gamma = \langle \alpha, \beta\rangle$, discrete group generated by $\alpha, \beta$ defined in eq.~\eqref{alpha and beta}.  In this case,
we introduce two chemical potentials $z_1$ and $z_2$ for  charges $J_3$ and $J_3’$,  for convenience.
$J_i , J'_i$ are generators of two $SU(2)$s
acting on two $\mathbb{C}^2$ factors in $\mathbb{C}^4$ respectively. That is
\begin{align}
J_3 = \frac{1}2 (h_1 + h_2)  = \textrm{diag} (\frac{1}2 , -\frac{1}2,0,0) \;, \nonumber
\\
J'_3 =\frac{1}2 (h_2 - h_1) = \textrm{diag}(0,0, \frac{1}2 , - \frac{1}2 )\; .
\end{align}
We use the normalization for $SU(2)$ generators such that $J_3$ has eigenvalues $\frac{1}2$ and $-\frac{1}2$ in minimal
(fundamental) representation.  The graviton index in $AdS_4 \times S^7$ can be written in terms of two
chemical potential $z_1 , z_2$ as follow
\begin{align}
I_{S^7;sp} (x, z_1 , z_2) &= I_{S^7;sp} (x, y_1 = \sqrt{z_1 z_2}, y_2 = \sqrt{z_1/ z_2} , y_3 = 1), \nonumber
\\
& = \sum  I_{S^7}^{(J, J')}(x) \chi^{SU(2)}_{J} (z_1 ) \chi^{SU(2)}_{J^\prime} (z_2).
\end{align}
Using the $SU(2)^2$ isometry in $S^7$ geometry, the index can be expanded in terms of two $SU(2)$ characters, $\chi_{J}(z_1 )$ and $\chi_{J'} (z_2)$. Recall that the discrete group $\Gamma$ is generated by two elements $\alpha = \exp (\frac{4\pi i J_3}{2m}) \otimes \mathbb{I}$ and  $\beta = \exp(\pi i J_2)\otimes \exp (2 \pi i J_3^\prime)$.  $\Gamma$-invariant states can be divided into following two types (depending on $J_3 =0 $ or $J_3 \neq 0 $),
\begin{align}
\textrm{A-type : } \; &|J, J_3 =0 \rangle \otimes |J^\prime, J_3^\prime \rangle \; \textrm{with $(-1)^{J +  2J'} =1 $}, \nonumber
\\
\textrm{B-type : } \; &(|J, J_3 \neq 0 \rangle + |J, - J_3\rangle ) \otimes |J^\prime, J_3^\prime\rangle \; \textrm{with $J_3 \in m \mathbb{Z}$ and $J'_3 \in \mathbb{Z}$}, \textrm{ or} \nonumber
\\
& (|J, J_3 \neq 0 \rangle - |J, - J_3\rangle ) \otimes |J', J_3'\rangle \; \textrm{with $J_3 \in m \mathbb{Z}$ and $J'_3 \in \mathbb{Z}+\frac{1}2$}.
\end{align}

Here, states are represented  by their total angular momentum $J$
and $z$-component  $J_z$ of two $SU(2)$. Other quantum numbers are
irrelevant and thus suppressed. It's easy to see that these states
are invariant under  $\alpha$. $\beta$-invariance of states in
A-type  can be shown as follows,
\begin{align}
\beta \cdot (|J, J_3 = 0 \rangle \otimes |J^\prime,J^\prime_3\rangle ) &= (-1)^{J} (-1)^{2J_3^\prime} (|J, J_3 = 0 \rangle \otimes |J',J'_3\rangle ) , \nonumber
\\
&= (-1)^{J+ 2J^\prime}(|J, J_3 = 0 \rangle \otimes |J',J'_3\rangle ) .
\end{align}
Here we use the fact that $(-1)^{\pi i J_2} |J_3 =0 \rangle = (-1)^{J} |J_3 =0 \rangle$ for $J \in \mathbb{Z}$ and $(-1)^{J_3^\prime} = (-1)^{J^\prime}$. To see the $\beta$-invariance of the states in B-type, one need to note that
\begin{align}
\{ \exp(\pi i J_2 ) , J_3 \} =0, \quad \textrm{when $J\in  \mathbb{Z}$}.
\end{align}
Thus $\beta$ flips the sign of quantum number $J_3$,
\begin{align}
\exp(\pi i J_2 ) \cdot |J, J_3\rangle = |J, -J_3 \rangle, \; \textrm{when $J \in \mathbb{Z}$ and $J_3 \neq 0$}.
\end{align}
Using this property one can easily check the $\beta$-invariance of
states in $B$-type. The index over gravitons in $AdS_4 \times
S^7/\langle \alpha, \beta \rangle = AdS_4 \times
(S^7/\mathbb{Z}_m)/\mathbb{Z}_2$ can be written as
\begin{equation}
I_{(S^7/\mathbb{Z}_m)/\mathbb{Z}_2:sp} (x ,y ) = I_{(S^7/\mathbb{Z}_m)/\mathbb{Z}_2:sp}^{(0)} (x)
 + \sum_{n>0} I^{(mn)}_{S^7;sp}(x)y^{mn}. \label{graviton  on S7/<alpha, beta>}
\end{equation}
Since $\beta$ flip  sign of  $J_3$ quantum number,  $J_3$ is no
longer a good quantum number in $S^7/ \langle \alpha, \beta
\rangle$. But $|J_3|$ is still a good quantum number and $y$ is
chemical potential for the quantum number. The quantum number
$|J_3|$ correspond to monopole charge $\sum_{i} |n_i|$ in
$\mathcal{N}=4$ $O(2N),USp(2N), SO(2N+1)$ SYMs which have $M$-theory
on $AdS_4\times S^7/\langle \alpha , \beta\rangle $ as gravity dual
in infrared limit. First term $I^{(0)}_{S^7/\langle \alpha, \beta
\rangle ;sp}$ collect index contribution form gravitons in  A-type,
that is
\begin{align}
I^{(0)}_{(S^7/\mathbb{Z}_m)/\mathbb{Z}_2:sp} (x)& = \sum_{J , J^\prime ; (-1)^{J+2J^\prime}=1}   (2J^\prime+1) I^{(J, J')}_{S^7} (x), \nonumber
\\
& = \frac{x (3+ 2 x^{1/2} + 2x - 2x^{5/2} -x^3)}{(1-x^2)^2}.
\label{I(0) on S7/<alpha,beta>}
\end{align}
The second term in \eqref{graviton  on S7/<alpha, beta>} comes from gravitons in B-type and
 $I^{(n)}_{S^7;sp}(x)$ here is same as that in \eqref{I^{(n)}_{S^7;sp}}.

\section{Large N index on $\mathcal{N}=4$ SYMs}

\subsection{$U(N) \oplus  ($m$ \textrm{ fundamentals}$)} \label{appB1}
Using the general superconformal index formula in section \ref{generalindex}, it straightforward to
write down  the superconformal index formula for the SYM theory.
\begin{align}
&I_{U(N): m}(x,y) = \sum_{\{ s=\{n_i\} \}} \frac{1}{\textrm{(sym)}} y^{m \sum_i n_i } x^{\epsilon_0 }
 \int \prod_{i=1}^N  d\lambda_i  \exp \big{[} \sum \frac{1}{n} f_{U(N)\oplus m} (x^n  , e^{i n \lambda_i} )\big{]},
 \nonumber
\\
&f_{U(N): m} (x, e^{i \lambda_i}) = 2 \sum_{i,j=1}^N \frac{x^{1/2}}{1+x }  e^{i (\lambda_i - \lambda_j ) }
 x^{|n_i - n_j|} + m \sum_{i=1}^N \frac{x^{1/2}}{1+x} (e^{i \lambda_i} + e^{- \lambda_i})x^{|n_i|} \nonumber
\\
& - \sum_{i \neq j} e^{i (\lambda_i - \lambda_j )}x^{|n_i - n_j|}, \quad \epsilon_0 = \frac{m}{2}\sum_i |n_i|.
\label{index formula U(N)+m}
\end{align}
Here the chemical potential $y$ for $U(1)_{diag}\subset U(N)$ monopole charge is introduced.
In the gravity side, the monopole charge quantum number can be identified with a generator
of $\mathbb{Z}_m$ in $AdS_4 \times S^7/\mathbb{Z}_m$.

To take  large N limit on the index, we  introduce distribution function $\rho(\theta)$ as
\begin{equation}
\rho(\theta) = \sum_{i=N_1 +1}^N \delta (\lambda_i - \theta) \;.
\end{equation}
Here $N_1$ denote the number of  non-zero monopole fluxes of $U(1)^N \subset U(N)$,
\begin{align}
s= \textrm{diag} \{n_1 , \ldots , n_{N_1} , 0, \ldots, 0 \}, \; \; n_i \neq 0. \label{U(N) monopole charge}
\end{align}
Since $\theta$ is periodic variable, it's convenient to introduce
Fourier transformation coefficients of $\rho(\theta)$, denoted by
$\{\rho_n\}$
\begin{align}
\rho_n = \int d\theta \rho(\theta)  e^{ in \theta} = \sum_{i=N_1 + 1}^N e^{i n \lambda_i}.
\end{align}
In the large N limit, holonomy integrals can be replaced by functional integral of distribution
fuction $\rho (\theta)$
\begin{align}
\int \prod_{i=1}^N d \lambda_i  \rightarrow  \int \prod_{i=1}^{N_1 } d\lambda_i \int D[\rho(\theta)] = \int  \prod_{i=1}^{N_1} d \lambda_i \int \prod_{n=1}^\infty d \rho_n d \rho_{-n}.
\end{align}
Using the variables $\{\rho_n\}$,  the index \eqref{index formula U(N)+m} for given monopole
charges \eqref{U(N) monopole charge} can be written as follows in the large N limit,
\begin{align}
&I_{U(\infty):m} (x, y)=  I_{N_1} (x,y)
\int \prod_{n=1}^\infty d^2 x_n \exp \big{[}\sum_{n=1}\frac{1}n \big{(}
 - \frac{1}2 x_n^T M (\cdot^n) x_n  + V^{T}(\cdot^n) x_n \big{)}\big{]}, \nonumber
\end{align}
where,
\begin{align}
&x_n := (\rho_n, \rho_{-n})^T, \; M = \left(
                                     \begin{array}{cc}
                                       0 & 1-2 h(x)  \\
                                       1- 2 h(x) & 0 \\
                                     \end{array}
                                   \right), \; h(x):= \frac{x^{1/2}}{1+x}\nonumber
\\
& V = \bigg{(}m h(x)+ [2 h(x)-1]\sum_{i=1}^{N_1} x^{|n_i|}e^{- i \lambda_i}, \; m h(x)
+ [2 h(x)-1]\sum_{i=1}^{N_1} x^{|n_i|}e^{ i \lambda_i} \bigg{)}^T.
\end{align}
$I_{N_1} (x,y)$ denote the index \eqref{index formula U(N)+m} with $N= N_1$ and monopole charge $ s_{N_1} = (n_1 , \ldots , n_{N_1})$. 
Performing the Gaussian integrations in $\vec{x}_n$ (ignoring $x$-independent factors),
\begin{align}
&\int \prod_{n=1}^\infty d^2 x_n  \exp \bigg{(}\sum_{n=1}\frac{1}n \big{(} - \frac{1}2 x_n^T M (\cdot^n) x_n  + V^{T}(\cdot^n) x_n \big{)}\bigg{)} \;, \nonumber
\\
&= \prod_{n=1} \frac{1}{\sqrt{\det M(\cdot^n)}} \exp \big{(}  \sum_{n=1}^\infty \frac{1}2 V(\cdot^n)^T M(\cdot^n)^{-1} V(\cdot^n) \big{)} \; , \nonumber
\\
&=\bigg{( }\frac{1}{\prod_{n=1}^\infty \big{[}1-2 h (x^n)\big{]}}  \exp \big{[} \sum_{n=1}^\infty \frac{1}n \frac{ m^2 h^2(x^n)}{1-2 h(x^n)} \big{]} \bigg{)}\times    \nonumber
\\
&\exp \bigg{(} \sum_{n=1}^\infty\frac{1}n  \big{(}-\sum_{i=1}^{N_1} m h(x^n) x^{n|n_i|}(e^{i n \lambda_i}+ e^{- i n \lambda_i})  + \sum_{i,j=1}^{N_1}  (1-2h(x^n)) x^{n(|n_i | + |n_j|)}e^{i n (\lambda_i -  \lambda_j)} \big{)} \bigg{)} \; .  \label{U(N) gaussian integration}
\end{align}
When all  monopole charges are zero, $N_1=0$ and the second factor in the above is 1 and the first factor give the large N index.  Thus the first factor can be considered as monopole zero index, which can be written as follows
\begin{align}
&I^{(0)}_{U(\infty):m}(x)  = \frac{1}{\prod_{n=1}^\infty \big{(}1-2 h (x^n)\big{)}}  \exp \big{(} \sum_{n=1}^\infty \frac{1}n \frac{ m^2 h^2(x^n)}{1-2 h(x^n)} \big{)}  =\exp\big{(} \sum_{n=1}^\infty \frac{1} n I^{(0)}_{U(\infty):m:sp} (x^n ) \big{)} \;, \nonumber
\\
&\textrm{with} \quad I^{(0)}_{U(\infty):m:sp} (x) =\frac{ 2 (x^{1/2} +  x -  x^{5/2}) }{ 1- x^{1/2} - x^2 + x^{5/2} } + (m^2 -1) \frac{x}{(1+x) (1- x^{1/2})^2  }.  \label{I^(0)_{U(inf)}}
\end{align}
Summarizing, the large N index is given by
\begin{align}
&I_{U(\infty):m}  (x, y) = I^{(0)}_{U(\infty):m}(x)   I^\prime_{U(\infty):m}’(x,y),  \; \textrm{where}\nonumber
\\
&I^\prime_{U(\infty):m}(x,y ) = I_{N_1} (x,y) \times (\textrm{second factor in eq.~\eqref{U(N) gaussian integration}})  \; , \nonumber
\\
&  =\frac{1}{\textrm{(sym)}}x^{\epsilon_0 }y^{ m \sum_{i=1}^{N_1} n_i} \int \prod_{i=1}^{N_1 }d \lambda_i  \exp \big{(}\sum_{n=1}^\infty \frac{1}n f^\prime_{U(\infty):m}(x^n, e^{i n  \lambda_i}) \big{)},  \; \textrm{with}\nonumber
\\
& f^\prime_{U(\infty):m}(x, y) = \sum_{i,j=1}^{N_1}  \bigg{(}  2 h(x) (x^{|n_i - n_j|} - x^{|n_i|+|n_j|})  - ( (1- \delta_{ij})x^{|n_i - n_j|} - x^{|n_i|+|n_j|}) \bigg{)} e^{i (\lambda_i - \lambda_j)}. \label{U(N) large N index}
\end{align}

As  with the large N index for ABJM theory \cite{Kim:2009wb}, the
large N index for the SYM exhibit following factorization properties
\begin{equation}
\sum_{\{n_i\}}I^\prime_{U(\infty):m}(x,y) = \big{(}\sum_{\{ n_i >0 \} } I^\prime_{U(\infty):m}(x,y)\big{)} \big{ (}\sum_{\{n_i <0 \}}I^\prime_{U(\infty):m} (x,y)\big{)} := I_{U(\infty):m}^{(+)} (x,y) I_{U(\infty):m}^{(-)}(x,y).
\end{equation}
From the large N index formula \eqref{U(N) large N index}, one can easily find following relation  between positive/negative monopole charge index $I^{(\pm)}_{U(\infty):m}$
\begin{equation}
I^{(+)}_{U(\infty):m} (x, y) = I^{(-)}_{U(\infty):m}(x,1/y). \label{relation I^{+} and I^{-}}
\end{equation}
Note that $m$ (number of fundamental hypermultiplet) dependence in the large N index $I^\prime_{U(\infty):m}$ only  appears  as pre-factor in front of holonomy integral,
\begin{align}
x^{\epsilon_0} y^{m \sum n_i}= x^{\frac{m}2 \sum |n_i|} y^{m \sum n_i} \; .
\end{align}
Due to this simple dependence of large N index on $m$,  one can easily relate $I_{U(\infty:m)}^{(\pm)}(x,y)$ for general $m$ to that  for $m=1$. Let $I^{(+)}_{U(\infty):m=1}(x,y)$ be written as
\begin{align}
&I_{U(\infty):m=1}^{(+)} (x,y) = \exp \big{[}\sum_{n=1}^\infty  \frac{1}n  I^{(+)}_{U(\infty):m=1:sp} (x,^n y^n)\big{]}, \nonumber
\\
&I^{(+)}_{U(\infty):m=1:sp} (x,y) = \sum_{n=1}^\infty I^{(n)}_{U(\infty):m=1:sp} (x) y^n.
\end{align}
Then for general $m$, the large N index $I^{(+)}_{U(\infty):m} (x,y)$ is given by
\begin{align}
&I_{U(\infty):m}^{(+)} (x,y) = \exp \big{[}\sum_{n=1}^\infty  \frac{1}n  I^{(+)}_{U(\infty):m:sp} (x,^n y^n)\big{]}, \; \textrm{with} \nonumber
\\
&I^{(+)}_{U(\infty):m:sp} (x,y) = \sum_{n=1}^\infty I^{(n)}_{U(\infty):m=1:sp} (x) x^{\frac{n(m-1)}  2} y^{mn}. \label{nonzero-monopole}
\end{align}
Negative monopole charge index $I^{(-)}_{U(\infty):m}$ also can be related to the case when $m=1$ using the
relation eq.~\ref{relation I^{+} and I^{-}}.

\subsection{O(2N), Sp(2N), SO(2N+1)} \label{largeN-bcd}
To write down the superconformal index for SO(2N)/Sp(2N)/SO(2N+1) gauge group, first we need to know
weights and root of these gauge group. Summarizing the results
\begin{align}
SO(2N) \qquad & \rho \in 2N \quad \{ \pm e_i \}_{i = 1, \cdots, N}  \nn \\
& \alpha \in G \quad \{ ( e_i - e_j ) \}_{i, j=1, \cdots, N}, \{ \pm ( e_i + e_j ) \}_{i<j = 1, \cdots, N} , \nn
\\ \nn
\\
Sp(2N)  \qquad & \rho \in 2N \quad \{ \pm e_i \}_{i = 1, \cdots, N} \nn \\
& \alpha \in G \quad \{ ( e_i - e_j)\}_{i,j=1,  \cdots, N}   , \{   \pm (e_i + e_j ) \} _{i<j=1, \cdots, N } , \{ \pm 2 e_i \}_{i=1, \cdots, N } \nn
\\ \nn
 \\
SO(2N+1 ) \qquad & \rho \in 2N+1 \quad \{ \pm e_i, 0\}_{i = 1, \cdots, N} \nn \\
& \alpha \in G \quad \{  (e_i - e_j ) \}_{i,j = 1, \cdots, N },  \{ \pm (e_i + e_j ) \}_{i<j = 1, \cdots, N} , \{  \pm e_i \}_{i = 1, \cdots, N}  \; . \nn \\
\nn
 \end{align}
In large N limit, $O(2N)$ SYM with $m$ fundamental hypermultiplets can't be distinguished from
the $SO(2N)$ SYM with same matter contents. Thus we will consider the large N limit on $SO(2N)$ SYM.
Using the general index formula in sec \ref{generalindex} and weight and roots of $SO(2N)$,
the superconformal index can be written as
\begin{align}
&I_{SO(2N):m}(x,y) = \sum_{\{s=\{n_i>0\}\}} \frac{1}{\textrm{(sym)}} y^{m \sum_i |n_i| } x^{\epsilon_0 } \int \prod_{i=1}^N  d\lambda_i  \exp \big{[} \sum \frac{1}{n} f_{SO(2N):m} (x^n  , e^{i n \lambda_i} )\big{]}, \nonumber
\\
&f_{SO(2N):m} (x, e^{i \lambda_i})  \nonumber
\\
&= \sum_{i,j=1}^N 2 h(x) e^{i (\lambda_i - \lambda_j)}x^{|n_i  - n_j|} +
\sum_{i<j}^N 2 h(x) ( e^{i(\lambda_i + \lambda_j)}  + e^{- i
(\lambda_i + \lambda_j)})  x^{|n_i + n_j|}, \; \; h(x):= \frac{x^{1/2}}{1+x}.\nonumber
\\
& + \sum_{i=1}^N 2 m h(x) (e^{i \lambda_i} + e^{- i
\lambda_i}) x^{|n_i|}- \sum_{i,j=1}^N e^{i (\lambda_i - \lambda_j)} x^{|n_i - n_j|} -
\sum_{i<j}^N (e^{i(\lambda_i + \lambda_j)} +e^{- i (\lambda_i
+ \lambda_j)} ) x^{|n_i + n_j|}. \label{index formula O(2N)+m}
\end{align}
We introduce the chemical potential $y$ for monopole charge $\sum_i |n_i|$. Using the $(\mathbb{Z}_2)^N$ symmetry in Weyl group of $O(2N)$ we can take all monopole charges to be positive.
Following the same procedure in the $U(N)$ SYM case, the large N
index of the $O(2N)$ SYM becomes
\begin{align}
&I_{O(\infty):m}(x, y)  \nonumber
\\
&= I_{O(N_1):m } (x,y) \int \prod_{n=1}^\infty d \chi_n \exp \bigg{(} - \sum_{n=1}^\infty \frac{1}{2n} \big{(}(1-2h(x^n))\chi_n^2 - (1-2 h(x^n))\chi_{2n}  \nonumber
\\
&- 4m h(x^n)\chi_n - 2\chi_n v_n \big{)} \bigg{)}, \quad  v_n := (2h(x^n)- 1) \sum_{i=1}^{N_1 } x^{n|n_i|}(e^{i n \lambda_i }+e^{- i n \lambda_i}).\nonumber
\\
&= I_{O(N_1):m } (x,y)\prod_{k=1}^\infty \frac{\exp [ \frac{ (2 m h (x^{2k-1})+ v_{2k-1})^2}{2(2k-1)(1-2h(x^{2k-1}))}+\frac{(1-2h(x^k)+2m h(x^{2k})+v_{2k})^2}{4k(1-2 h(x^{2k}))}]}{{\sqrt{1-2 h(x^{2k-1})}}\sqrt{1-2 h(x^{2k})}}\;, \nonumber
\\
&:= I_{O(\infty):m}^{(0)}(x) I^\prime_{O(\infty):m}(x,y )\; . \label{largeNofO(N)}
\end{align}
Here we introduce variables $\chi_n = \sum_{i=N_1}^N  \big{(}  e^{i \lambda_i n } + e^{- i \lambda_i n } \big{)}$, with monopole charge $s=\{ n_1, \ldots, n_{N_1},0,\ldots, 0 \}$.
Monopole charge zero sector  index $I^{(0)}_{O(\infty):m}(x)$ is given by
\begin{align}
& I^{(0)}_{O(\infty):m} (x) \nonumber
\\
&= \frac{1}{\prod_{n=1}^\infty \sqrt{1- 2 h(x^n)}} \exp \big{(}\sum_{n=1}^\infty \frac{1}n  (\frac{2m^2 h^2}{1- 2h} (x^n)  + \frac{m h(x^{2n}) (1- 2
h (x^n))+ (h(x^n)-\frac{1}2)^2 }{1- 2h(x^{2n})})
\big{)} \nonumber
\\
&= \exp \big{[} \sum_{n=1}^\infty \frac{1}n I^{(0)}_{O(\infty):m:sp}(x^n)\big{]}  \;, \; \textrm{where}\nonumber
\\
&I^{(0)}_{O(\infty):m:sp}(x)=
\frac{x(3+2x^{1/2} + 2x - 2x^{5/2} -x^3)}{(1-x^2)^2} \nonumber \\
\quad & +(2m^2 +m) \frac{x}{(1-x)^2}+(2m^2 - m -1) \frac{ 2 x^{3/2}}{ (1-x)^2 (1+x)}
\label{no-monopole-so2n}
\end{align}
The remaining part in large N index $I^\prime_{O(\infty):m}(x,y)$ in
eq.~\eqref{largeNofO(N)} is given by
\begin{align}
&I^\prime_{O(\infty):m}(x,y)    \nonumber
\\
&=I_{O(N_1):m } (x,y)\prod_{k=1}^\infty \exp \bigg{(} \frac{4 m h (x^{2k-1}) v_{2k-1} + v_{2k-1}^2}{2 (2k-1)(1-2 h (x^{2k-1}))} + \frac{(4m h (x^{2k})+ 2(1- 2h (x^k)))v_{2k} + v_{2k}^2}{4k (1- 2h (x^{2k}))}\bigg{)} \nonumber
\\
& =  \frac{1}{(\textrm{sym})} y^{m \sum_{i=1}^{N_1}|n_i|} x^{\epsilon_0} \prod_{i=1}^{N_1} d\lambda_i  \exp \big{(} \sum_{n=1}^\infty \frac{1}n  f^{\prime}_{O(\infty):m} (x^n, e^{i n \lambda_i})\big{)}\;, \;\; \textrm{with} \nonumber
\\
&f^{\prime}_{O(\infty):m} = \sum_{i,j=1}^{N_1}  \bigg{(}  2 h(x) (x^{|n_i - n_j|} - x^{|n_i|+|n_j|})  - ( (1- \delta_{ij})x^{|n_i - n_j|} - x^{|n_i|+|n_j|}) \bigg{)} e^{i (\lambda_i - \lambda_j)}.
\end{align}

Note that $I^\prime_{O(\infty):m}(x,y)$ is same with positive monopole charge part $I^{(+)}_{U(\infty):m}(x, y)$ for $U(N)$ SYM case \eqref{U(N) large N index}. This is consistent with the fact that
non-zero KK momentum graviton index on $AdS_4 \times (S^7/\mathbb{Z}_m)/\mathbb{Z}_2$ is equal to positive KK momentum graviton index on $AdS_4 \times S^7$ , which is  explicitly shown in eq.~\eqref{graviton  on S7/<alpha, beta>}.
\\
Using the same technique, one can calculate the large N index for
$SO(2N+1)$, $Sp(2N)$ SYM with $m$ fundamental hypermultiplets. For
$SO(2N+1)$ SYM, the large N index is same as $O(2N)$ SYM. For
$Sp(2N)$ SYM, the non-zero monopole charge sector of large N index
coincide with $O(2N)$ SYM case, that is
\begin{align}
I^{\prime}_{Sp(\infty):m }(x,y)  = I^{\prime}_{O(\infty):m} (x,y) \;.
\end{align}
But in zero monopole charge sector large N index is somewhat different, the result is
\begin{align}
&I^{(0)}_{Sp(\infty):m:sp}(x) = I^{(0)}_{O(\infty):-m:sp}(x) \nonumber
\\
&=
\frac{x(3+2x^{1/2} + 2x - 2x^{5/2} -x^3)}{(1-x^2)^2}  +(2m^2 -m) \frac{x}{(1-x)^2}+(2m^2 + m -1) \frac{ 2 x^{3/2}}{ (1-x)^2 (1+x)} \;  . \label{no-monopole-sp2n}
\end{align}

\section{Partition Function}

Let us consider the brane configurations given in fig.~\ref{fig4}. In (a), there are $N$ D3-branes diagramed
by the circle, $n$ NS5-branes, and $ m = m_1 + m_2 + \cdots + m_L$ D5-branes, $m_i$ D5-branes between
the $i$th and $i+1$th NS5-branes. As in section 5, we take T-duality to obtain SCS theories from brane configuration in (b).

The fig.~\ref{fig4} (a) corresponds to 3-d, ${ \cal N}=4$, $U(N)^n$ SYM with $n$ bi-fundamental, and $m$ fundamental hyper-multiplets in following representation,
\begin{align}
&(N, \bar{N}, 1, \ldots, 1) \oplus (1, N, \bar{N}, 1, \ldots, 1) \oplus \ldots \oplus ( \bar{N}, 1, \ldots, 1, N) \nn \\
&  m_1 ( N, 1, \cdots, 1) \oplus m_2 ( 1, N, 1, \cdots, 1)  \oplus \cdots \oplus m_{L} ( \underbrace{1, \cdots, 1}_{L-1}, N, \underbrace{ 1, \cdots, 1}_{n-L}) \nn
\end{align}
The IR limit of fig.~\ref{fig4} (b) gives ${ \cal N}=4$, $U(N)^{n+m}$ CS with $(n+m)$ bi-fundamental hyper-multiplets in
\be  (N, \bar{N}, 1, \cdots, 1) \oplus (1, N, \bar{N}, 1, \cdots, 1) \cdots \oplus ( \bar{N}, 1, \cdots, 1, N), \nn
\ee
 The Chern-Simons level can be one of $ 1,-1, 0$ , depending on whether D3-branes are between NS$^{ \prime}$/D5, D5/NS$^{\prime}$, (D5/D5 or NS$^{\prime}$/NS$^{\prime}$).  The Chern-Simons levels from the brane configuration (b) can be denoted as a following $(n+m)$-vector,
\begin{align}
&
  \vec{k} = ( 1, \underbrace{0, \cdots,   0}_{m_1 -1 } , -1, 1,  \underbrace{0,  \cdots, 0}_{m_2 -1},  -1, \cdots, \underbrace{0, \cdots, 0}_{m_L -1} , -1, \underbrace{0, \cdots, 0}_{n-L}  ) \label{cs-level}
 \end{align}
 Note that for the same number of 5-branes, the moduli spaces are same \cite{Imamura:2008nn} though the corresponding theories can differ by the ordering of the 5-branes.

  We will show that the partition function for two theories are same. The procedure is similar to the proof in \cite{Kapustin:2010xq} which shows the equivalence of the partition function of $U(N)_1 \times U(N)_{-1}$ ABJM and the 3-d ${ \cal N}=8$, $U(N)$ YM. However we will show the calculation explicitly to be self-contained.
 The partition function of ${ \cal N}=4$ SCS is given by
 \be
 Z_{CS}  = \frac{1}{(N!)^{n+m} } \int ( d^N \sigma)^{n+m} \frac{ \prod_{ i < j} \sinh^2 ( \pi  \sigma^1_{ ij} ) \cdots \sinh^2 ( \pi \sigma_{ ij}^{ n+m} ) }{ \prod_{ i,j} \cosh ( \pi ( \sigma_i^1 - \sigma_j^2 ) ) \cdots \cosh ( \pi ( \sigma_i^{n+m} - \sigma_j^1 ) )  } e^{i S_0 [ \sigma]}
 \label{zcs}
 \ee
 where $ \sigma^I_{i}$ for $ I = 1, \cdots, n+m$, $i = 1, \cdots, N$ are the moduli of the theory.
 $S_0$ is the classical action,
 \be S_0 =  \pi  \sum_{i=1}^N [ ( \sigma^1_i)^2 - ( \sigma^{m_1 +1 }_i)^2 + \cdots  \cdots - ( \sigma_i^{m+L})^2 ]  = \pi \sum_{i=1}^N \sum_{I=1}^{n+m} k_I ( \sigma^I_i)^2
 \nn
 \ee
 where $k_I$ is the Chern-Simons level of $I$th gauge group, given in \eqref{cs-level}.
 Let us define a sequence $ \{n \}$ as $ (n_1, n_2, \cdots, n_L ) : = ( m_1+1, m_1+m_2+2, \cdots, (m_1+ m_2 + \cdots + m_L )+L)$ to denote the gauge groups with $k = -1$, and define $ \{ \tilde{n} \}$ as $ ( \tilde{n}_1, \cdots, \tilde{n}_L) : = (1, m_1+2, m_1+m_2+3, \cdots, (m_1 + \cdots + m_{L-1})+ L ) $  to denote the gauge groups with $k=1$. The classical action can be written as $S_0 = \pi    \sum_{I = 1}^L  \sum_i ( ( \sigma_i^{ \tilde{n}_I})^2 - ( \sigma_i^{n_I})^2 ) $. Introduce permutations $ \rho_I : \{1, \cdots, N\} \to  \{ 1, \cdots, N \}$, then the partition function becomes
  \begin{align}
Z_{CS}
 &= \sum_{ \rho_1, \cdots, \rho_{n+m} }  \frac{ (-1)^{ \rho_1 + \cdots + \rho_{n+m} } }{(N!)^{n+m}} \int   ( d^N \sigma )^{n+m}  \frac{e^{ \pi i   \sum_{I = 1}^L ( ( \sigma^{ \tilde{n}_I})^2 - ( \sigma^{n_I})^2 )}  }{  \prod_{ i} \cosh ( \pi ( \sigma_i^{ 1}- \sigma_{\rho_1(i)}^{2})  )   \cdots \cosh ( \pi ( \sigma^{n+m}_i - \sigma^1_{\rho_{n+m}(i) } ) }.
\nn \end{align}
using \eqref{fourier}.
One can redefine variables $ \sigma^2_{ \rho_1 (i)} \to \sigma^2_i$, repeatedly for $ \sigma^I_{ \rho_{I-1} (i)}$, then for $ \rho: = \rho_1 + \rho_2 + \cdots + \rho_{n+m}$, it becomes
\begin{align}
Z_{CS}
 &= \sum_{ \rho }  \frac{ (-1)^{ \rho } }{N!} \int   ( d^N \sigma )^{n+m}  \frac{e^{ \pi i   \sum_{I = 1}^L ( ( \sigma^{ \tilde{n}_I})^2 - ( \sigma^{n_I})^2 )}  }{  \prod_{ i} \cosh ( \pi ( \sigma_i^{ 1}- \sigma_{i}^{2})  ) \cosh ( \pi ( \sigma_i^{2}- \sigma_{i} ^{3} ) )   \cdots \cosh ( \pi ( \sigma^{n+m}_i - \sigma^1_{\rho(i) } ) }  .
\nn \end{align}
Now use \eqref{fourier-2} to introduce new integral variables $ \tau^I_i$. Then integrate over the variables of the gauge groups with vanishing Chern-Simons levels using  \eqref{fourier-3}
to obtain
\begin{align}
&   \sum_{ \rho }  \frac{ (-1)^{ \rho } }{N!}  \int ( d^N \sigma \  d^N \tau )^{n+m}  \frac{e^{ \pi i \sum_i  \sum_{ I=1}^L ( ( \sigma_i^{n_I })^2 - (\sigma_i^{ \tilde{n}_I})^2  )}  e^{ 2 \pi i \sum_i \sum_I ( \tau_i^I (\sigma_i^{I}- \sigma_i^{I+1}) )  }   }{ \prod_i \cosh ( \pi \tau_i^1 ) \cosh ( \pi \tau_i^2 )   \cdots \cosh ( \pi \tau_i^{n+m} ) }
\nn \\
& =\sum_{ \rho }  \frac{ (-1)^{ \rho } }{N!}    \int (d^N \sigma  \ d^N \tau )^{2L}
 \frac{e^{ \pi i  \sum_{i,I} ( ( \sigma_i^{n_I })^2 - (\sigma_i^{ \tilde{n}_I})^2  )}   e^{ 2 \pi i ( \sigma_i^{ \tilde{n}_1}   ( \tau_i^{ \tilde{n}_1}- \tau_{ \rho(i)}^{ n_L } )+ \cdots+ \sigma_i^{ \tilde{n}_L }  ( \tau_i^{ \tilde{n}_L}- \tau_i^{ n_{L-1} }  )  + \sigma_i^{ n_L} ( \tau_i^{n_L}- \tau_i^{  \tilde{n}_{L}} )   )  }   }{  \prod_i \cosh ( \pi \tau_i^{n_{1}} ) \cdots \cosh ( \pi \tau_i^{n_{L-1} } )       \cosh^{n-L + 1} ( \pi \tau_i^{n_L} )  \prod_{i,I}   \cosh^{m_{I}} ( \pi \tau_i^{ \tilde{n}_I})  } \ .
\nn \end{align}
Doing the Gaussian integral results in
\begin{align}
Z_{CS}& = \sum_{ \rho}  \frac{ (-1)^{ \rho}}{ N!}  \int ( d^N \tau)^{2L}
 \frac{  e^{2  \pi i \sum_i \left( \tau^{ \tilde{n}_1} ( \tau^{n_1}_i - \tau^{n_L}_{ \rho(i)} ) +  \sum_{I=2}^L  \tau_i^{ \tilde{n}_I}  ( \tau_i^{n_I} - \tau_i^{n_{I-1} } ) \right) }    }{ \prod_i (\prod_{I=1}^{L-1} \cosh ( \pi \tau_i^{n_{I}} )  ) \cosh^{n-L + 1} ( \pi \tau_i^{n_L} )  \prod_{I = 1}^L \cosh^{m_{ I}}( \pi \tau_i^{ \tilde{n}_I} )   } \ .  \label{z-cs}
\end{align}

 On the other hand,  the partition function of the ${ \cal N}=4$, $U(N)^n$ YM  is given by
 \begin{align}
 Z_{ YM} & = \frac{1}{(N!)^n} \int (d^N \sigma)^n \frac{ \prod_{ i<j} \sinh^2 ( \pi \sigma_{ ij}^1 ) \cdots \sinh^2 ( \pi \sigma_{ ij}^n ) }{ \prod_{ i, j} \cosh ( \pi ( \sigma_{i}^1 - \sigma_j^2 ) ) \cdots \cosh ( \pi ( \sigma_i^n - \sigma_j^1 ) ) \prod_{i=1}^N  \prod_I \cosh^{m_I} ( \pi \sigma_i^I )  }
 \label{zym}  \end{align}
 One can now use \eqref{fourier} to rewrite $\frac{ \prod_{i<j} \sinh( \pi \sigma^I_{ij} ) \sinh ( \pi \sigma^{I+1}_{ij} )}{  \prod_{i,j}  \cosh ( \pi ( \sigma^I_i - \sigma^{I+1}_j) ) }= \sum_{ \rho_I} (-1)^{ \rho_I} \frac{1}{ \cosh ( \pi ( \sigma^I_i - \sigma^{I+1}_{\rho_I (i)}))} $, then redefine variables repeatedly to get
 \begin{align}
 Z_{YM}
& = \sum_{ \rho} \frac{ (-1)^{ \rho} }{ N! }  \int ( d^N \sigma)^{ n }  \frac{1}{   \prod_i \left(  \left[ \prod_{I=1}^{n-1} \cosh ( \pi ( \sigma_i^{ I} - \sigma_i^{ I+1} )) \right]   \cosh ( \pi ( \sigma_i^n - \sigma_{ \rho(i)}^1 ) )   \prod_{ I=1}^n \cosh^{m_{ I} } ( \pi \sigma_i^{ I} ) \right)
 } .
 \nn \end{align}
 Now use \eqref{fourier-2} to introduce $\tau^I_i$ variables, then integrate out $ \sigma^{L+1}_i, \cdots, \sigma^{n}_i$ using \eqref{fourier-3} to obtain
 \begin{align}
Z_{YM}
 & = \sum_{ \rho} \frac{ (-1)^{ \rho}}{N!}  \int ( d^N \sigma)^L (d^N \tau)^L  \frac{ e^{ 2 \pi i \sum_i ( \sigma_i^1 (\tau_i^{1}- \tau_{ \rho(i)}^{L}) + \sigma_i^2 ( \tau_i^{2}- \tau^{1}_i ) + \cdots + \sigma_i^L ( \tau_i^{L}- \tau_i^{ L-1} ) ) } }{ \prod_i  \left(  \left( \prod_{I=1}^{L-1} \cosh ( \pi \tau_i^I ) \right)  \cosh^{n-L+1} ( \pi \tau_i^L ) \prod_{ I = 1}^L \cosh^{m_{ I} } ( \pi \sigma_i^{ I} ) \right)  } \ .  \nn
 \end{align}
It is equivalent to the integral in \eqref{z-cs}. This shows that  for any 5-brane configuration of fig.~\ref{fig4} type,  the partition functions of ${ \cal N}=4$ CS's  and YM's are same
 \begin{align}
 Z_{CS} = Z_{YM} .  \nn \end{align}

\subsection{Useful Formulas} \label{useful}
In \cite{Kapustin:2010xq}, the following identity is proved
\be
\frac{ \prod_{i<j} \sinh ( x_i - x_j) \sinh (y_i - y_j)}{ \prod_{i,j} \cosh ( x_i - y_j) } = \sum_{ \rho} (-1)^{ \rho} \prod_i \frac{1}{ \cosh ( x_i - y_{ \rho (i)} )} , \label{fourier}
\ee
 where $(-1)^{ \rho}$ is defined to be $1 (-1)$ for an even(odd) permutation $ \rho$.

The following identities are useful forms of Fourier transform,
\begin{align} \frac{1}{ \cosh( \pi \sigma) } & = \int d \tau \frac{e^{ 2 \pi i \tau \sigma}}{ \cosh ( \pi \tau) },  \label{fourier-2} \\
 \delta ( \tau^I - \tau^J) & = \int d \sigma e^{ 2 \pi i ( \tau^I - \tau^J ) \sigma }. \label{fourier-3}
\end{align}

\providecommand{\href}[2]{#2}

\begingroup\raggedright

\end{document}